\documentclass[journal,onecolumn,12pt]{IEEEtran}
\ifCLASSINFOpdf
\else
\fi
\usepackage{amsmath}
\usepackage{url}
\usepackage{booktabs} 
\usepackage[utf8]{inputenc}
\usepackage{multirow}
\usepackage{graphicx}
\usepackage{enumitem}
\usepackage{subfloat}
\usepackage{verbatim}
\usepackage{cite}
\usepackage{amsmath,amssymb,amsfonts}
\usepackage{algorithm}
\usepackage{algorithmic}
\usepackage{textcomp}
\usepackage{diagbox}
\usepackage{amsmath}
\usepackage{color} 
\usepackage{fancyhdr}
\usepackage{caption}
\usepackage{subcaption}
\usepackage[dvipsnames]{xcolor}
\usepackage{epstopdf}

\DeclareMathOperator*{\argmax}{arg\,max}
\DeclareMathOperator*{\argmin}{arg\,min}

\begin{document}
	%
	\title{Adversarial Attacks and Defense Methods for Power Quality Recognition}
	%
	%
	%
	
	\author{Jiwei~Tian, Buhong~Wang,~\IEEEmembership{Member,~IEEE}, Jing~Li, Zhen Wang and Mete~Ozay
		\thanks{An early version of this work was presented in part at the 2019 IEEE SmartGridComm \cite{tian2019adaptive}. This work was supported by the National Natural Science Foundation of China under Grant 61902426 and the National University of Defense Technology Research Fund under Grant ZK21-44. Jiwei Tian is with Air Traffic Control and Navigation College, Air Force Engineering University, Xi’an, China (e-mail: tianjiwei2016@163.com). B. Wang and Z. Wang are with Information and Navigation College, Air Force Engineering University, Xi’an, China. J. Li is with Henan University of Technology, Zhengzhou, China. }}
	
	%
	%

	\markboth{Technical Report}%
	{Shell \MakeLowercase{\textit{et al.}}: Bare Demo of IEEEtran.cls for IEEE Journals}
	%



	\maketitle
	
	\begin{abstract}
		Vulnerability of various machine learning methods to adversarial examples has been recently explored in the literature. Power systems which use these vulnerable methods face a huge threat against adversarial examples. To this end, we first propose a  signal-specific method and a universal signal-agnostic method to attack power systems using generated adversarial examples. Black-box attacks based on transferable characteristics and the above two methods are also proposed and evaluated. We then adopt adversarial training to defend systems against adversarial attacks. Experimental analyses demonstrate that our signal-specific attack method provides less perturbation compared to the FGSM (Fast Gradient Sign Method), and our signal-agnostic attack method can generate perturbations fooling most natural signals with high probability. What's more, the attack method based on the universal signal-agnostic algorithm has a higher transfer rate of black-box attacks than the attack method based on the signal-specific algorithm. In addition, the results show that the proposed adversarial training improves robustness of power systems to adversarial examples.
	\end{abstract}
	
	\begin{IEEEkeywords}
		Adversarial example, adversarial training, universal perturbation.
	\end{IEEEkeywords}

	%
	\IEEEpeerreviewmaketitle

	\section{Introduction}
	%
	%
	%
	%
	\IEEEPARstart{P}{ower} quality refers to a  variety of electromagnetic phenomena that characterize voltage and current measured at a given time instance and location in a power system \cite{5154067}. Disturbance of power quality (PQ) signals can cause severe problems in  electrical grids \cite{igual2018integral}. PQ disturbances occurring in  grids may cause financial loss,  damage and faulty operation of equipment installed in grids and end-user devices \cite{mahela2015critical}. In end-user systems, PQ disturbances can cause loss of data, memory failures of sensitive loads (such as computers, protection and relay devices), and unstable operation of  controls \cite{deokar2014integrated}. Therefore, if PQ disturbances are not properly recognized and mitigated by monitoring systems, then power grid assets can be harmed resulting in high costs and safety hazards in extreme but realistic cases.
	Numerous techniques have been proposed to address the aforementioned problems \cite{ mahela2015critical,mishra2018power}. Among them, deep neural networks (DNNs) have been employed to improve accuracy for recognition of PQ disturbances replacing hand-crafted features by learned feature representations \cite{wang2019novel}.
	

	
	Although DNNs show outstanding performance in various tasks, they are vulnerable to a form of attacks called  ``adversarial example" \cite{goodfellow2014explaining}. Adversarial examples with different attributes have been explored for  image, text, and speech processing in  \cite{goodfellow2014explaining,xu2019adversarial,8846746}. DNNs have been recently used in cyber physical systems (CPSs) \cite{fei2019cps}, but exploration and assessment of their security is still an open problem. Security risks caused by adversarial machine learning (ML) methods in CPSs have been discussed  in  \cite{bor2019adversarial}. Vulnerabilities of ML algorithms utilized in power systems were first shown in \cite{B1}. Additional works have addressed security problems for grid event classification \cite{niazazari2019adversarial}, N-1 security  \cite{venzke2019verification}, power quality signal classification \cite{B1} and detection of false data injection attacks  \cite{saygheevasion,saygheadversarial}. The vulnerabilities of load forecasting algorithms caused by  perturbations injected into input temperatures were studied in \cite{chen2019exploiting}. A domain-specific framework was proposed to evaluate security and resilience of load forecasting algorithms  in \cite {zhou2019evaluating} and experimental results showed that DNN-based load predictions may suffer from worst-case attacks even with a partial control of network. From a defense perspective, adversarial training \cite {tang2019enabling} and game theory \cite {barreto2019design} are respectively adopted to design resilient load forecasting systems. Adversarial attacks for DNN-based state estimation were studied in \cite{liu2019adversarial}. 
	
	Characterization of these attacks and defense mechanisms is an open problem that has not been fully explored and understood. An adversarial signal attack is a vital threat, since signals generated by the attack are perceptually similar to the original signals. Therefore, a human operator of the system may not be aware of the presence of the attack. Then, the system may take a wrong action according to the incorrectly predicted signal class affecting the physical system. Vulnerability of DNNs against particular adversarial attacks in power systems was analyzed in \cite{B1,tian2019adaptive}. Among them, a variant of FGSM attack \cite{B14} is proposed to generate adversarial examples for PQ classification in \cite{B1}. A more accurate and computationally efficient attack method and corresponding defense method are proposed in \cite{tian2019adaptive}. However, stronger and more practical attack methods (such as as universal perturbations and black-box attacks) and corresponding defense strategies need to be explored more comprehensively, which is lacking in the current literature.
	
	In order to defend power systems against these attacks and improve their robustness,  learning models deployed in the systems should be carefully analyzed. In this paper, we explore adversarial attacks to CNN (Convolutional Neural Network) algorithms and examine the corresponding defense methods for PQ recognition tasks. The main contributions of this work are summarized as follows\footnote{The code used for reproducing the experimental analyses and results are provided on \url{https://github.com/JiweiTian/AT_SG.git}.}:
	\begin{itemize}
		\item We propose a signal-specific algorithm which outperforms a state-of-the-art method \cite{B1} for generating adversarial perturbation of signals in power systems.
		\item We analyze existence of universal signal-agnostic perturbations generated by CNNs in power systems, and propose an algorithm to recognize such perturbations. Our results show that generated universal perturbations make models discriminate natural power signals\footnote{We consider  power signals commonly measured by a particular system operator of a power system as natural power signals.} into several statistically dominant signal categories.
		\item  We explore and analyze black-box attacks: the attacker
			can train a local substitute model and generate adversarial
			samples based on the substitute model, and then use its
			transferable characteristics to attack the target model. The
			experimental results show that the attack method based
			on the universal signal-agnostic algorithm has a higher
			transfer rate of black-box attacks than the attack method
			based on the signal-specific algorithm, which is a huge and practical threat for CNN based PQ classification.
		\item We propose using adversarial training as a defense method to improve robustness of learning models employed in power systems to adversarial attacks.
		\item We quantitatively analyze misclassification of different attacks, and summarize the characteristics of different attack methods for PQ signals. We suggest an approach to qualitatively analyze the relationship between the characteristics of adversarial attacks to PQ signals and defense methods in the Smart Grid. For this purpose,  we visualize signal measurements and their feature representations learned at different layers of CNNs using the t-SNE algorithm \cite{maaten2008visualizing}.
	\end{itemize}

	The paper is organized as follows. In Section II, we provide an overview of adversarial methods and describe our proposed attack algorithms. In Section III, our proposed adversarial training method is introduced. Experimental analyses of the proposed methods  are given for various case studies in Section IV. Conclusion and discussion are provided in Section V.

	\section{Adversarial Attacks for Power Quality Classification using CNNs}
	\begin{figure}[t]
		\centerline{\includegraphics[scale=0.2]{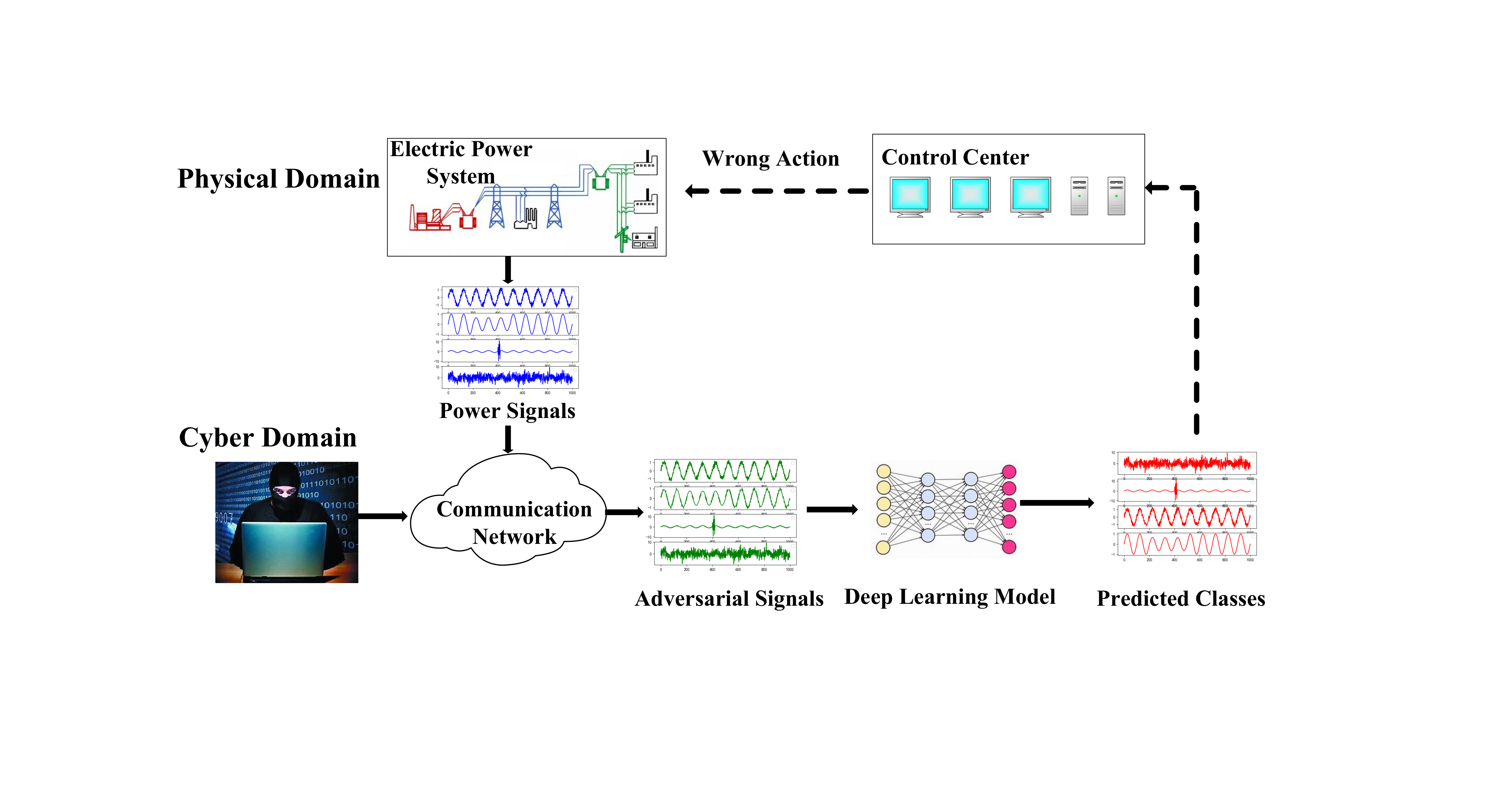}}
		\caption{Adversarial attacks on power signal recognition algorithms using deep learning methods in the Smart Grid.}
		\label{PowerAIattack}
	\end{figure}
	\subsection{Attack Models and Objectives}

	As shown in Fig.~\ref{PowerAIattack}, a well-trained CNN can be used for
	power quality assessment in the Smart Grid, which is crucial
	for its robust operation. However, an attacker can penetrate
	a communication network of the system, and use adversarial
	signals instead of original signals to attack the system to fool
	the CNN for misclassification of the PQ signals (for example, Man-In-The-Middle attack). Note that
	we consider attacks in the testing phase (adversarial examples and adversarial attacks) rather than in the training phase
	(poisoning attacks). The objective of the attacker is generating
	imperceptible adversarial signals similar to the original signals,
	resulting in their misclassification by algorithms without being
	noticed by human operators\footnote{Suppose that the original signal belongs to Class 1. Since the added perturbation is so small (imperceptible), it should belong to the same type as the original signal from the human operator’s point of view (Class 1). However, the target CNN will regard it as a new type (for example, Class 2)}.
	
	In our  white-box attack model, we assume that attackers can access  parameters of CNNs deployed in power systems and use this knowledge to construct adversarial signals. In practice, these parameters can be obtained by insider threats or using various attack methods, such as eavesdropping on network traffic or database intrusion. Rouhani et al. \cite{8677311} suggest that white-box attacks represent the most powerful attacker that can appear in real-world settings. Although these assumptions render a strong attacker that may not always represent the
		practical cases, they enable us to evaluate the robustness and
		vulnerabilities under the worst-case scenario, providing an
		upper bound on the impact of proposed attacks on PQ CNN
		models.
	
	Although many adversarial attack methods have been proposed in image domain, there are significant differences between PQ recognition tasks using power signals and object recognition tasks using two or three dimensional images:
\begin{itemize}
	\item \textit{Data Characteristics:} A PQ signal is a time-series, whose characteristics are different from that of two or three dimensional images.  Therefore, in order to produce imperceptible PQ perturbations, the characteristics of time-series data should be considered to avoid occurrence of abnormal values with great changes. Besides,  although image pixels take values from a fixed interval, such as [0,1] or [0,255], PQ signals have no definite range. 
	
	\item \textit{Levels of Perturbation:} In image processing and computer vision tasks, norms such as $\ell_0$, $\ell_2$ and $\ell_{\infty}$ are used to model adversarial perturbations. Since PQ signals have no definite range, some adversarial attack methods based on box constraints \cite{B13} and  ``change of variables" such as C$\&W$ \cite{carlini2017towards} are not suitable for PQ signals. Although  $\ell_0$ norm-based perturbations such as Sparsefool \cite{modas2019sparsefool} and ``One pixel attack" \cite{su2019one} in image domain will not affect human perception, the outliers generated by the $\ell_0$ norm-based perturbations will have a greater impact on time-series PQ signals. Besides, for PQ signals, it's hard to determine a maximum budget for $\ell_{\infty}$ norm-based perturbations (the maximum budget will affect human perception of PQ signals, and each value is allowed to be changed by up to this limit, with no limit on the number of input data that are modified). Therefore, we consider the $\ell_2$ norm case for PQ signals in the paper.

		\end{itemize}
	
Although FGSM was employed in particular power systems\footnote{Therefore, we consider the FGSM method as the baseline and state-of-the-art for adversarial attacks in PQ recognition in our experimental analyses.} \cite{B1}, development of a general and suitable approach for adapting adversarial attacks for PQ recognition in power systems is an open problem. We address this problem by proposing two methods called signal-specific and signal-agnostic  attacks. We realize these approaches to generate adversarial signals for PQ recognition by adapting DeepFool \cite{B5} and universal attack methods in power systems. Although other adversarial attack methods have been proposed in image domain, many methods and their variants are not suitable for the PQ signal field for various reasons, such as the existence of box constraints (L-BFGS \cite{B13}, C$\&W$ \cite{carlini2017towards}) and artificially determined attack parameters (FGSM \cite{B14}, I-FGSM \cite{kurakin2016adversarial}, JSMA \cite{papernot2016limitations}). We consider that our proposed approaches can be employed to adapt a wider class of attack models for various tasks in power systems by researchers, especially attack methods based on geometrical perspective \cite{fawzi2017robustness} (such as targeted DeepFool \cite{liao2018backdoor}\cite{esmaeilpour2020sound}, DeepFool-5 \cite{lu2017safetynet} and so on).

	\subsection{Signal-Specific Adversarial Signals}
	In this section, we introduce our signal-specific adversarial attack method which is proposed to generate adversarial perturbations of  signals in power systems. 
	
	Suppose that $f: \mathbb{R}^n \to \mathbb{R}^K$ is a classifier computed by a learning model to determine the class label of a signal $\boldsymbol{x}_{0} \in \mathbb{R}^n$ among $K$ power quality classes.
	In this work, deep learning algorithms are used to  compute feature representations of PQ signals and classifiers $f$. Therefore, we will identify the classifier by a non-linear function $f(\boldsymbol{x}_{0}:\mathcal{W})$ which is parameterized by a set of tensors $\mathcal{W}$ denoting weights and bias vectors called model parameters of CNNs.
	The parameters $\mathcal{W}$ are optimized using a stochastic optimization algorithm such as stochastic gradient descent. Since the target
		model in this paper is fixed, for the sake of simplicity, we
		don’t show the dependence of $f$ on $\mathcal{W}$. The class label of the power signal $\boldsymbol{x}_{0}$ is estimated by 
	\begin{equation}
	\hat{y}\left(\boldsymbol{x}_{0}\right) = \argmax \limits_k f_k(\boldsymbol{x}_{0}),
	\label{eq:opt1}
	\end{equation}
	where $f_k(\boldsymbol{x}_{0})$ is the output of $f(\boldsymbol{x}_{0})$  for the $k^{th}$ class. An adversarial perturbation is defined as the minimal (imperceptible) perturbation $\boldsymbol{r}$ that is sufficient to change the estimated power quality label $\hat{y} \left(\boldsymbol{x}_{0}\right)$ at  $\boldsymbol{x}_{0}$ by
	\begin{equation}
	\label{eq:opt2}
	\begin{aligned}
	\Delta(\boldsymbol{x}_{0};\hat{y}(\boldsymbol{x}_{0})) &\triangleq \min _{\boldsymbol{r}} \| \boldsymbol{r} \|_2  \\
	{\rm subject \; to} \; \quad \hat{y}(\boldsymbol{x}_{0}& + \boldsymbol{r}) \neq  \hat{y}(\boldsymbol{x}_{0}),
	\end{aligned}
	\end{equation}
	where $\| \cdot \|_2$ denotes the $\ell_2$ norm. 
	
	In adversarial attacks, attackers can make learned models incorrectly estimate class labels of attacked power signals by adding the minimal imperceptible perturbation computed by \eqref{eq:opt2}. For instance, normal power signals can be misclassified as impulse signals, and vice versa. Thereby, adversarial attacks can affect power quality monitoring systems by adversarial data injection, and cause failure of their operation. The minimal perturbation computed using \eqref{eq:opt2} for different classes and samples of PQ signals may be different from each other. To generate minimum perturbation of the signal  ${\boldsymbol{x}_{0}}$ optimizing the closest decision boundary of $f$ from $\boldsymbol{x}_{0}$, a signal-specific adversarial attack method (SSA) is proposed in Algorithm \ref{alg:deepfool}.

	CNNs partition high dimensional class regions with nonlinear decision boundaries by locally piecewise linear functions implemented in network nodes into subregions \cite{linear} which are convex polytopes \cite{expressive}.  Wong and Zolter \cite{prov} propose that if CNNs are optimized to minimize the worst case loss over convex outer approximation of the set of activations reachable through a norm-bounded perturbation (called adversarial polytope), then the trained CNN models is provably robust to any norm-bounded adversarial attack. We employ this result to generate adversarial signals according to a convex outer approximation of an adversarial polytope of trained CNN models. For this purpose, we first define the region (polytope) of the space of adversarial signals by
	\begin{equation}
	\mathcal{P} \triangleq \bigcap \limits _{k=1} ^K \{ \boldsymbol{x}: f_{\hat{y}\left(\boldsymbol{x}_{0}\right)}\left(\boldsymbol{x}_{0}\right) \geq f_{k}\left(\boldsymbol{x}\right) \}.
	\end{equation}
	Activations of the model for the signals $\boldsymbol{x}$ belonging to this region for each different $k^{th}$ class are bounded by that for the signals $\boldsymbol{x}_{0}$ for the predicted class $\hat{y}\left(\boldsymbol{x}_{0}\right)$.  
	Next, we generate adversarial signals in Algorithm 1 as follows:
\begin{itemize}
	    \item \textbf{Step 4 and 5:} The region $\mathcal{P}$ is approximated at each $i^{th}$ epoch by
	\begin{equation}
	\hat{\mathcal{P}}_i = \bigcap \limits _{k=1} ^K \{ \boldsymbol{x}: f_{k}^{'} +  (\boldsymbol{w}_{k}^{\prime}) ^{T} \boldsymbol{x} \leq 0 \},
	\end{equation}
	where $\boldsymbol{w}_{k}^{\prime} = \nabla f_{k}\left(\boldsymbol{x}_i \right) ^T  - \nabla  f_{\hat{y}\left(\boldsymbol{x}_{0}\right)}\left(\boldsymbol{x}_i \right) ^T  $ (at step 4) and $f_k' = f_{k}\left(\boldsymbol{x}_i\right)  - f_{\hat{y}\left(\boldsymbol{x}_{0}\right)}\left(\boldsymbol{x}_{i}\right)$ (at step~5). $\boldsymbol{w}_{k}^{\prime}$ denotes the difference between gradients of functions $f$ of $\boldsymbol{x}_i$ for the $k^{th}$ class and for the predicted class $\hat{y}\left(\boldsymbol{x}_{0}\right)$ with respect to model parameters.
	
	\item \textbf{Step 7:} We compute a convex outer approximation (boundary) of the polytope and generate signals far from this boundary. Therefore, first the closest hyperplane  $\hat{l}$ of the boundary of $\mathcal{P}$ is computed where $|\cdot|$ denotes the absolute value. 
	
	\item \textbf{Step 8:} Next, the minimum perturbation $\boldsymbol{r}_{*}\left(\boldsymbol{x}_{0}\right)$ projecting $\boldsymbol{x}_{0}$ on $\hat{l}$ is approximated at each $i^{th}$ epoch by $\boldsymbol{r}_i$. 
	
	\item \textbf{Step 9 and 10:} Then, the adversarial signal $\boldsymbol{x}_{i}$ and epoch variable $i$ are updated at the step 9 and 10, respectively. 
	
	\item \textbf{The steps 3-10} are iterated while estimated class labels of the normal  $\boldsymbol{x}_{0}$ and the adversarial signal $\boldsymbol{x}_{i}$ are same.
	
	\item \textbf{Step 12:}
	Finally, the algorithm injects the sum of perturbations $\boldsymbol{r}_{j}, j = 1,2, \ldots, i$ to the original signal to compute the adversarial PQ signal.

\end{itemize}

	\begin{algorithm}[t]
		\caption{Generating Signal-Specific Adversarial Signals.} 
		\label{alg:deepfool} 
		\begin{algorithmic}[1] 
			\REQUIRE 
			$\boldsymbol{x}$: An input power quality signal, \\
			
			$f$: A learned signal classification model. \\
			\ENSURE
			Adversarial power quality signal $\hat{\boldsymbol{x}}$.
			\STATE {Initialization: 
				$i \leftarrow 0$ and $\boldsymbol{x}_{0} \leftarrow \boldsymbol{x}$}.
			\WHILE {$\hat{y}\left(\boldsymbol{x}_{i}\right)=\hat{y}\left(\boldsymbol{x}_{0}\right)$}
			\FOR {$k \neq \hat{y}\left(\boldsymbol{x}_{0}\right)$}
			\STATE {$\boldsymbol{w}_{k}^{\prime} \leftarrow \nabla f_{k}\left(\boldsymbol{x}_{i}\right)-\nabla f_{\hat{y}\left(\boldsymbol{x}_{0}\right)}\left(\boldsymbol{x}_{i}\right)$}
			\STATE {$f_{k}^{\prime} \leftarrow f_{k}\left(\boldsymbol{x}_{i}\right)-f_{\hat{y}\left(\boldsymbol{x}_{0}\right)}\left(\boldsymbol{x}_{i}\right)$}
			\ENDFOR
			\STATE {$\hat{l} \leftarrow \argmin \limits _{k \neq \hat{y}\left(\boldsymbol{x}_{0}\right)} \frac{\left|f_{k}^{\prime}\right|}{\left\|\boldsymbol{w}_{k}^{\prime}\right\|_{2}}$}
			\STATE {$\boldsymbol{r}_{i} \leftarrow \frac{\left|f_{\hat{l}}^{\prime}\right|}{\left\|\boldsymbol{w}_{\hat{\imath}}^{\prime}\right\|_{2}^{2}} \boldsymbol{w}_{\hat{l}}^{\prime}$}
			\STATE {$\boldsymbol{x}_{i+1} \leftarrow \boldsymbol{x}_{i}+\boldsymbol{r}_{i}$}
			\STATE {$i \leftarrow i+1$}
			\ENDWHILE
			\RETURN Adversarial power quality signal $\boldsymbol{x}+\sum \limits _{j=1} ^i \boldsymbol{r}_{j}$.
		\end{algorithmic} 
	\end{algorithm}
	
	\subsection{Signal-Agnostic Adversarial Signals}
	In our proposed signal-agnostic adversarial attack method (SAA), attackers aim to generate signal-agnostic and very small perturbation vectors that cause normal PQ signals to be misclassified with \textit{high probability}.  An attacker can make the PQ signal be misclassified with high probability by adding signal-agnostic perturbations without spending time calculating the specific perturbation, independent of class of actual PQ signal (e.g., normal, sag and so on) fed to the system. Presence of such perturbations poses a great threat to learning models in power systems, as the adversary may simply add the same pre-computed signal-agnostic perturbation to the new signal and cause misclassification. Therefore, pre-computed signal-agnostic perturbations are called universal perturbations \cite{moosavi2017universal}.
	
	The goal of the SSA is to generate a quasi-imperceptible universal perturbation signal vector $\boldsymbol{v}$ such that \textit{most} PQ signals sampled from a distribution $\mu$ are misclassified when applying the universal perturbation signal. 
	We compute the vector $\boldsymbol{v}$ considering the following two constraints:
	\begin{enumerate}
		\item The magnitude of the universal perturbation vector $\boldsymbol{v}$ is controlled by a parameter $\xi >0$ using $\lVert \boldsymbol{v} \rVert_2 \le \xi.$
		
		\item The misclassification probability is lower bounded  for all signals $\boldsymbol{x}$ sampled from the distribution $\mu$ by
		\begin{equation}
		\underset{\boldsymbol{x} \sim \mu}{P}(\hat{y}(\boldsymbol{x} + \boldsymbol{v}) \neq  \hat{y}(\boldsymbol{x})) \ge 1-\delta , {\rm where} \; \delta \in (0,1).
		\end{equation}
	\end{enumerate}

	Let $\mathcal{X}=\left\{ \boldsymbol{x}_i \right\}_{i=1}^m$  be a set of PQ signals sampled from the distribution $\mu$. Our method aims at finding a perturbation $\boldsymbol{v}$ fooling signals satisfying $\lVert \boldsymbol{v} \rVert_2 \le \xi$. Our proposed algorithm (Algorithm \ref{alg:universal}) gradually crafts adversarial perturbations on signals belonging to $\mathcal{X}$. At each iteration, the minimal perturbation $\Delta \boldsymbol{v}_i$ pushing the sample $\boldsymbol{x}_i$ to the decision boundary is calculated and added to the current perturbation. More precisely, if the current perturbation $\boldsymbol{v}$ does not fool the sample signal $\boldsymbol{x}_i$, then we will compute an additional perturbation $\Delta \boldsymbol{v}_i$ with minimal norm to fool the sample signal $\boldsymbol{x}_i$  at step 5 of the Algorithm \ref{alg:universal} by solving the following optimization problem
	\begin{equation}
	\begin{aligned}
	\Delta \boldsymbol{v}_i  & =  \argmin \limits _{\boldsymbol{r}}\,\,\lVert \boldsymbol{r} \rVert _2 \\ 
	{\rm subject \; to}  &  \quad \hat{y}\left( \boldsymbol{x}_i+\boldsymbol{v}+\boldsymbol{r} \right) \ne \hat{y}\left( \boldsymbol{x}_i \right). 
	\label{eq:pert}
	\end{aligned}
	\end{equation}
	
	In order to meet the first constraint, the universal perturbation is projected on an $\ell_2$ ball of radius $\xi$ and centered at 0 at step 6 of the Algorithm \ref{alg:universal} by
	\begin{equation}
	\begin{aligned}
	\Pi_{2,\xi}\left( \boldsymbol{v} + \Delta \boldsymbol{v}_i \right)  & =  \argmin \limits _{\boldsymbol{v}'} \lVert \boldsymbol{v} + \Delta \boldsymbol{v}_i -\boldsymbol{v}' \rVert _2  \\
	{\rm subject \; to}  & \quad   \lVert \boldsymbol{v}' \rVert_2 \le \xi.
	\end{aligned}
	\end{equation}

	The fooling rate is defined for the perturbed training set $\mathcal{X}$ by
	\begin{equation}
	R\left( \mathcal{X},\boldsymbol{v} \right) :=\frac{1}{\sigma({\mathcal{X}})} \sum_{\boldsymbol{x} \in {\mathcal{X}}}{\mathbb{I}({\hat{y}\left( \boldsymbol{x}_i+\boldsymbol{v} \right) \ne \hat{y}\left( \boldsymbol{x}_i \right)}} ),
	\end{equation}
	where $\mathbb{I}(\cdot)$ is an indicator function and $\sigma({\mathcal{X}})$ denotes the size of the perturbed training set $\mathcal{X}$. If the rate $R\left( \mathcal{X},\boldsymbol{v} \right)$ exceeds $1-\delta$, then the algorithm terminates. This method does not ensure that we can find universal perturbations with large misclassification rates. However, in our experiments, we can obtain universal perturbations with relatively high misclassification rates using this method with limited signals.
	
	\begin{algorithm}[t]
		\caption{Generating Signal-Agnostic Adversarial Signals.} 
		\label{alg:universal} 
		\begin{algorithmic}[1] 
			\REQUIRE 
			$\mathcal{X}$: A set of power quality signals. \\
			$\xi$: An upper bound of the $\ell_{2}$ norm of the perturbation. \\
			$\delta$: A parameter used to control  the fooling rate. \\
			$f$: A learned signal classification model.
			\\
			\ENSURE
			A signal-agnostic perturbation vector $\boldsymbol{v}$.
			\STATE {Initialization: $\boldsymbol{v} \leftarrow \boldsymbol{0}$  }
			\WHILE {$R\left( \mathcal{X},\boldsymbol{v} \right) \le 1-\delta $}
			\FOR {each signal $\boldsymbol{x}_i\in \mathcal{X}$}
			\IF {$\hat{y}(\boldsymbol{x} + \boldsymbol{v}) =  \hat{y}(\boldsymbol{x})$}  
			\STATE {Compute the minimal perturbation $\Delta \boldsymbol{v}_i$ that
				pushes ${\boldsymbol{x}_i + \boldsymbol{v}}$ to the decision boundary by solving \eqref{eq:pert}.} 
			\STATE {Update the perturbation by}
			{$\boldsymbol{v}\gets \Pi_{2,\xi}\left( \boldsymbol{v}+\Delta \boldsymbol{v}_i \right)$.}
			\ENDIF
			\ENDFOR
			\ENDWHILE
			\RETURN Signal-agnostic perturbation vector $\boldsymbol{v}$.
		\end{algorithmic} 
	\end{algorithm}

	\subsection{Black-box Attacks Against CNN Based Power Quality Recognition}
	Without knowing the target model $f$, the attacker can train his own substitute model $f^{'}$ and craft adversarial examples against $f^{'}$. Due to the transferability of adversarial examples \cite{papernot2017practical,papernot2016transferability}, this kind of black-box attacks also has a certain success rate. If the adversarial example generated against the model $f^{'}$ can also mislead the model $f$, we call this black-box attack successful.
	
	Given the well-trained substitute model $f^{'}$, the attacker can use various attack
	methods to generate adversarial examples. In other words,
	the attacker can carry out white-box attacks based on the substitute
	model $f^{'}$ to generate corresponding adversarial examples, then
	use them to implement black-box attacks against the target
	model $f$. In the following experiments, we will explore and
	analyze black-box attacks based on the substitute model $f^{'}$ and the above proposed algorithms (Algorithm 1 and Algorithm
	2).
	
	\section{Defending Learning Models using Adversarial Training}
	Adversarial training is used to improve robustness of a learning model to adversarial attacks. In this approach, a model is trained using a large number of adversarial examples generated by different attack methods. The basic requirement of this approach is employment of the strongest possible attack to generate adversarial examples  using \eqref{eq:opt1} and \eqref{eq:opt2}. Adversarial training improves robustness of DNNs to adversarial attacks \cite{B14,B17}, provide regularization \cite{B14} for DNNs and improve their classification accuracy \cite{B16}.
	In this paper, we  adopt adversarial training as our defense method to improve robustness of PQ classification models in power systems. Our proposed adversarial training method is summarized in Algorithm \ref{alg:advtrain}.

	\begin{algorithm}[t]
		\caption{Defending Models using Adversarial Training.} 
		\label{alg:advtrain} 
		\begin{algorithmic}[1] 
			\REQUIRE 
			$D_{tr}$: A training dataset of clean samples. \\
			$N_{iter}$: The number of iterations used for training.   
			\ENSURE A robust power quality classification model $f_{\Theta}$, where $\Theta$ is a set of learnable parameters of the model $f$. \\	
			\textbf{Training the model $f_{\Theta}$ using a training dataset $D_{tr}$:}
			\FOR {$1,2, \ldots, N_{iter}$}
			\STATE {Optimize $\Theta$ by training $f_{\Theta}$ using   $D_{tr}$}.
			\ENDFOR
			\\
			\STATE{Generate a set of adversarial signals $\tilde{\mathcal{X}}$ based on normal signals $\mathcal{X}$ using Algorithm~1.}  \\
			
			\textbf{Retraining the model $f_{\Theta}$ using the set of signals $\tilde{\mathcal{X}}$:}
			\FOR {$1, 2, \ldots, N_{iter}$}
			\STATE {Update $\Theta$ by training $f_{\Theta}$ using  $\tilde{\mathcal{X}}$ and $\mathcal{X}$.}
			\ENDFOR
			\RETURN A robust power quality classification model $f_\Theta$.
		\end{algorithmic} 
	\end{algorithm}
	
	\begin{table}[t]
		\caption{Comparison of the FGSM, SSA and SAA.}
		\centering
		\scalebox{1.5}{\begin{tabular}{|c|c|c|c|}
			\hline
			& $\hat{\rho}_{\mathrm{adv}}$ & Misclassification Rate & Average Time (sec.) \\ \hline
			FGSM & $770\times 10^{-3}$               & 0.88                       & $1.2\times 10^{-2}$            \\ \hline
			SSA  & $5.0\times 10^{-3}$              & 0.98                       & $2.2\times 10^{-1}$            \\ \hline
			SAA  & $26.0\times 10^{-3}$              & 0.74                       & $1.6\times 10^{-6}$            \\ \hline
		\end{tabular}}
		\label{FGSM vs SSA}
	\end{table}
	
	\section{Experimental Analyses}
	\label{sec:exp}
	\subsection{Experimental Setup}
\begin{figure*}[htbp]
		\centerline{\includegraphics[scale=0.5]{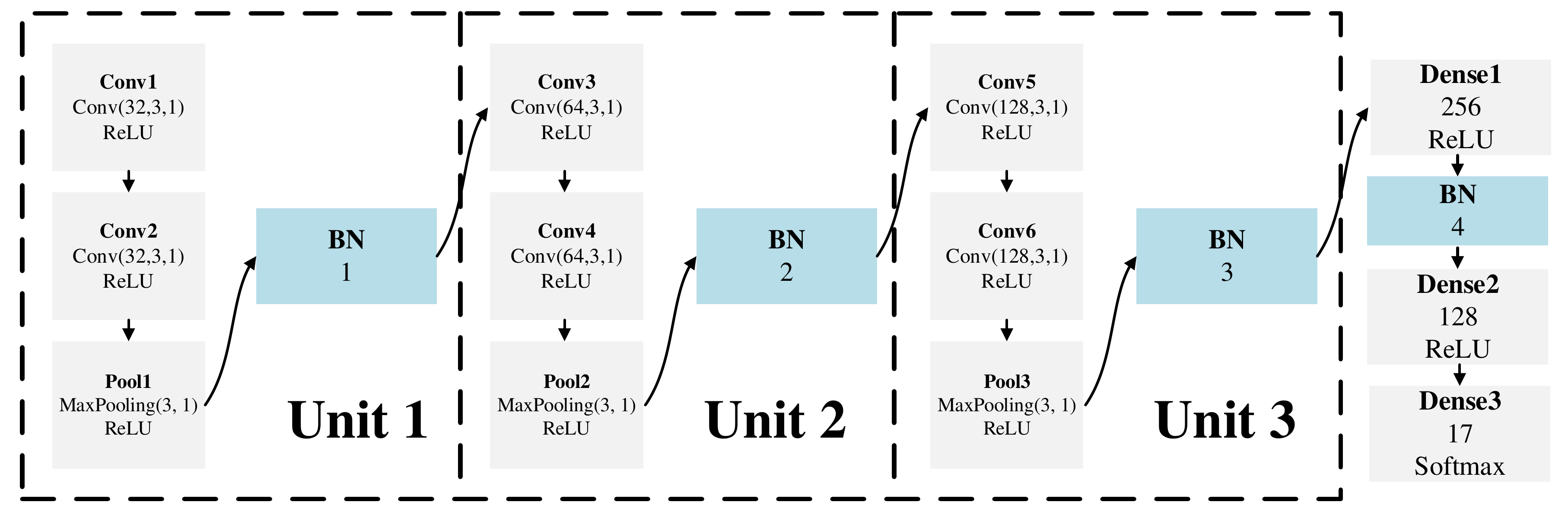}}
		\caption{Architecture of the CNN used in the experiments, where Conv(number of features, kernel size, stride) denotes a convolution layer, MaxPool(kernel size, stride) denotes a pooling layer and BN denotes batch normalization.}
		\label{CNNmodel}
	\end{figure*}

	We evaluate accuracy of the proposed attack and defense methods for power quality assessment by classifying signals using a convolutional neural network (CNN) \cite{wang2019novel}. Based on the PQ models in \cite{wang2019novel}, the sampling frequency of signals is set as 3200Hz, the fundamental frequency of  signals is set as 50Hz, the number of total cycles (periods) of the fundamental frequency is set as 10, and the amplitude (per unit) is set as 1. Therefore, the dimension of input signal vectors is 640. The network architecture is given in Fig.~\ref{CNNmodel}. 
	In the PQ assessment task, a CNN model was trained to classify signals into 17 categories. Signal types and corresponding representative symbols are shown in Table \ref{type}. We use the mathematical model of PQ disturbances and parameters proposed in \cite{igual2018integral}. A labeled dataset of 255000 signals at 30dB of SNR (Signal-Noise-Ratio) was constructed using 15000 signals belonging to each class. The dataset was randomly shuffled, 1/4 of the shuffled data was split as the testing set, and the rest was used for training. 
	\begin{table}[htbp]
		\caption{Signal types and corresponding representative symbols.}
		\centering

		\scalebox{1.25}{
		\begin{tabular}{|c|c|c|c|c|c|}
			\hline
			C-1 & Normal                  & C-7  & Harmonics            & C-13 & Sag with Oscillatory transient   \\ \hline
			C-2 & Sag                     & C-8  & Harmonics with Sag   & C-14 & Swell with Oscillatory transient \\ \hline
			C-3 & Swell                   & C-9  & Harmonics with Swell & C-15 & Sag with Harmonics               \\ \hline
			C-4 & Interruption            & C-10 & Flicker              & C-16 & Swell with Harmonics             \\ \hline
			C-5 & Transient/Impulse/Spike & C-11 & Flicker with Sag     & C-17 & Notch                            \\ \hline
			C-6 & Oscillatory transient   & C-12 & Flicker with Swell   &      &                                  \\ \hline
		\end{tabular}}
	\label{type}
	\end{table}

	In order to analyze accuracy of a learning model using clean signals, we randomly shuffled the dataset, and trained and tested the model 10 times. We trained each model for 15 epochs and Adam \cite{kingma2014adam} was used as the optimizer ($\alpha=0.001$, $\beta_1=0.9$, $\beta_2=0.999$, $\epsilon=1e^{-8}$). 
	We obtained $92.05\% \pm 0.11$ and $92.01\% \pm 0.11$ training and test accuracy, respectively. One trained model with $98.46\%$ test accuracy is used as our attack target model. For SSA, we generated corresponding adversarial signals for all clean signals including training data and test data. For SAA, we used a subset of training signals to generate universal perturbations.
	
	\begin{figure*}[t]
		\centerline{\includegraphics[scale=0.47]{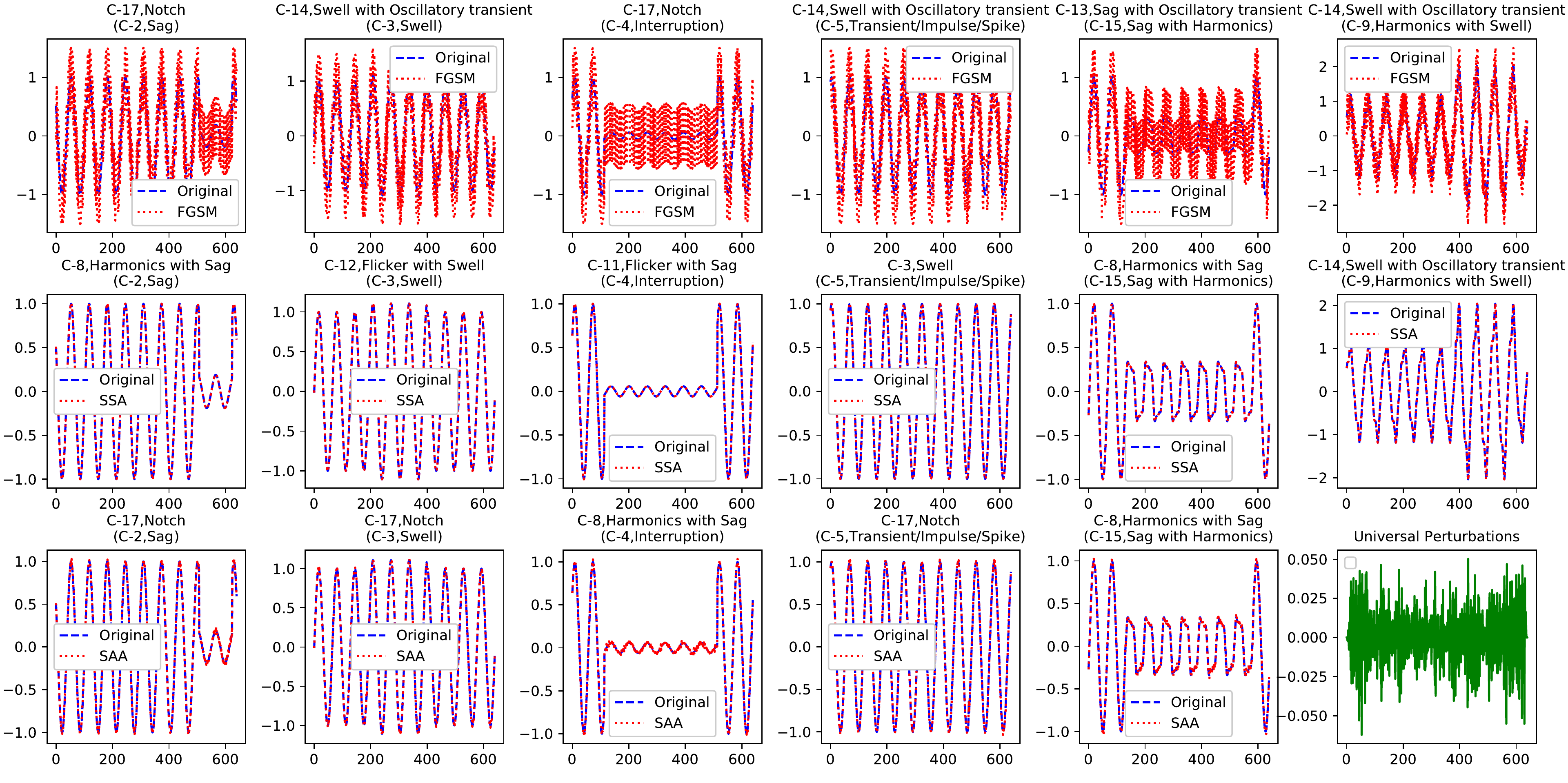}}
		\caption{Experimental analysis of classification of randomly chosen clean signals (original) and the corresponding adversarial signals generated using the FGSM (first row), SSA (second row) and SAA (third row) methods. In titles of subfigures, C-$i$ (C-$j$) denotes a case where a signal belongs to class C-$j$ while its class label is predicted as C-$i$. The x-axis denotes time (s) and the y-axis denotes Voltage/Current (V/A p.u.).}
		\label{Universal}
	\end{figure*}
	
	
	\subsection{Analysis of Signal-Specific Adversarial Signals}
	
	We compare  performance of our proposed SSA and FGSM \cite{B14} for generating signal-specific adversarial signals. We evaluate their performance using average robustness \cite{B5} by
	\begin{equation}
	\hat{\rho}_{\mathrm{adv}}=\frac{1}{\sigma({D})} \sum_{\boldsymbol{x} \in {D}} \frac{\|\hat{\boldsymbol{r}}(\boldsymbol{x})\|_{2}}{\|\boldsymbol{x}\|_{2}},
	\end{equation}
	where $\hat{\boldsymbol{r}}(\boldsymbol{x})$ is the perturbation vector generated using an attack method (e.g., FGSM or SSA), and $\sigma(D)$ denotes the size of the test set $D$. As shown in Fig.~\ref{Universal}, magnitudes of the adversarial perturbations generated by the SSA are very small and the adversarial signals are imperceptible to human operators. The $\epsilon$ parameter of the FGSM \cite{B14} is set as 0.5 to achieve the overall 0.88 misclassification rate which is proportion of signals whose labels are changed when perturbations are applied. According to the characteristics of FGSM attack, the larger the parameter $\epsilon$, the larger the adversarial perturbations, and the higher the corresponding misclassification rate. Besides, the selection of the parameter $\epsilon$ has no effect on the running time. We choose $\epsilon=0.5$ to achieve roughly the same misclassification rate as the other two methods, so as to analyze the effects of different attack methods. In Fig.~\ref{Universal}, we observe that magnitudes of the adversarial perturbations generated by the FGSM \cite{B14} are larger and more perceptible than those generated by the SSA. 
	
	In Table \ref{FGSM vs SSA}, we compare misclassification rate, average robustness and average running time of FGSM \cite{B14} and SSA. Table~\ref{FGSM vs SSA} shows that the SSA achieves a higher misclassification rate with smaller perturbation compared with the FGSM. Besides, we notice that the average running time spent for adversarial generation using the SSA is larger than using the FGSM. For example, the average running time spent for generating one adversarial signal is 0.22 seconds for the SSA, while that for the FGSM is 0.012 seconds. SSA spent most of the time for optimization of perturbation using Algorithm~\ref{alg:deepfool}. 

	\begin{table*}[htbp]
		\caption{Misclassification rate on the test set. Note that although  universal perturbations are computed on small sets of signals, the misclassification rate on test set is large.}
		\centering
		\scalebox{1.2}{
		\begin{tabular}{|c|c|c|c|c|}
			\hline
			Number of signals in training set (ratio) & 5000 (1.96$\%$)& 10000 (3.92$\%$)& 15000 (5.88$\%$)& 20000 (7.84$\%$)\\ \hline
			Misclassification rate ($\%$)       & 28   & 35    & 52    & 74    \\ \hline
		\end{tabular}
		\label{bar_SAA}}
	\end{table*}

	\subsection{Analysis of Signal-Agnostic Adversarial Signals}
	
	Unlike FGSM and SSA, SAA computes signal-agnostic (universal) perturbations using a training set of signals. We observed that the average time spent for training SAA to achieve misclassification rate 0.74 is 30 minutes. Less than 10$\%$ of the total signals were used to generate the universal perturbation. Table ~\ref{bar_SAA} shows that one can build relatively generalizable
	universal perturbations with few signals. Besides, additional time is not spent for optimization during testing, and the universal perturbation is applied to each test signal in real-time ($1.6\times 10^{-6}$ sec.).   
	Universal perturbations and adversarial samples generated using $\xi=1$ are shown in Fig.~\ref{Universal}  (the average $l_2$ norm of a signal is 18). In the experiments, the misclassification rate can reach to 0.74 and most natural signals are misclassified with high probability.

	Fig.~\ref{Universal} shows that voltage values of the universal perturbations are smaller and more imperceptible to human operators compared to the perturbations generated by FGSM. Although SSA and SAA can generate perceptually similar adversarial samples for some classes, the samples generated by different methods can be misclassified to different classes. For instance, class labels of the adversarial samples  belonging to $C$-3 which are generated by SSA and SAA, are incorrectly predicted by the learned model as $C$-12 and $C$-17, respectively. 
	

	\subsection{Analysis of Performance of Black-Box Attacks}
	
	For training substitute models, we suppose that the attacker
	can access some data in the training set (obtained by invading a
	database or building an experimental platform for simulation).
	Two types of black-box setting are considered: 
	\begin{itemize}
	\item Type I: The
	attacker knows the network architecture of the CNN model and does not know the specific model parameters. 
	\item Type II: Both network architecture and specific parameters of the CNN model are not known to attackers (the substitute model architecture has several layers more or less than the corresponding target model in our experiments).
	\end{itemize}
	
	Besides, we also consider datasets of different sizes used for training substitute models, that is, using $1/20$, $1/10$ and $1/5$ of the entire training set. For each black-box type and training data ratio, we trained 10 substitute models independently. The average test accuracy values of well-trained substitute models are shown in Table \ref{bbb}. Based on the well-trained substitute models, we generated adversarial signals based on the proposed attack methods in Section II. In detail, for each well-trained substitute model, we generated corresponding adversarial signals against the
	entire test data.  Then, we use these generated adversarial signals to attack the target model. 
	
	Since two attack methods (SSA and SAA) are proposed in Section II, we use these two methods independently to generate adversarial signals and evaluate their effects. The average misclassification rates of adversarial signals against the target model are shown in Table \ref{ccc} and \ref{ddd}, respectively. Table \ref{ccc} shows that black-box attacks for SSA have a low success rate. Besides, the average misclassification rate of Type I is higher than that of Type II, which implies that the substitute model using the same architecture with the target model has a higher success rate. In contrast, Table \ref{ddd} implies that black box attacks for SAA have a high success rate, and there is no significant difference in the misclassification rate under the two black-box types. This interesting phenomenon reflects that in black-box attacks, signal-agnostic (universal) perturbations have a higher misclassification rate than signal-specific perturbations. The above experimental results show that the attacker can train his substitute models and generate corresponding universal perturbations to carry out black-box attacks, which is a huge and practical threat for CNN based PQ classification.
	
	\begin{table}[ht]
		\centering

		\caption{Average test accuracies of well-trained substitute models.}
		\scalebox{1.5}{
		\begin{tabular}{|c|c|c|c|}
			\hline
	Training data ratio          & $1/20$ &  $1/10$ & $1/5$  \\ \hline
			Type I    & 94.9$\%$     & 96.4$\%$     & 99.3$\%$    \\ \hline
			Type II & 93.9$\%$  & 97.3$\%$  & 94.4$\%$    \\ \hline
		\end{tabular}}
		\label{bbb}
	\end{table}

	\begin{table}[ht]
		\centering
		\caption{Average misclassification rate of black-box attacks for SSA.}
		\scalebox{1.5}{\begin{tabular}{|c|c|c|c|}
			\hline
			Training data ratio      & $1/20$ &  $1/10$ & $1/5$  \\ \hline
			Type I    & 4.7$\%$     & 5.1$\%$     & $5.3\%$    \\ \hline
			Type II & 3.5$\%$  & 3.6$\%$  & 4.1$\%$    \\ \hline
		\end{tabular}}
		\label{ccc}
	\end{table}

	\begin{table}[ht]
		\centering
			\caption{Average misclassification rate of black-box attacks for SAA.}
			\scalebox{1.5}{\begin{tabular}{|c|c|c|c|}
				\hline
				Training data ratio          & $1/20$ &  $1/10$ & $1/5$  \\ \hline
				Type I    & 73.2$\%$     & 74.1$\%$     & 75.6$\%$    \\ \hline
				Type II & 72.6$\%$  & 75.5$\%$  & 76.1$\%$    \\ \hline
			\end{tabular}}
			\label{ddd}
	\end{table}
	
	\subsection{Graphical Analysis of Class Predictions under Attacks}
\begin{figure}[H]
	\centering
	\centerline{\includegraphics[scale=0.45]{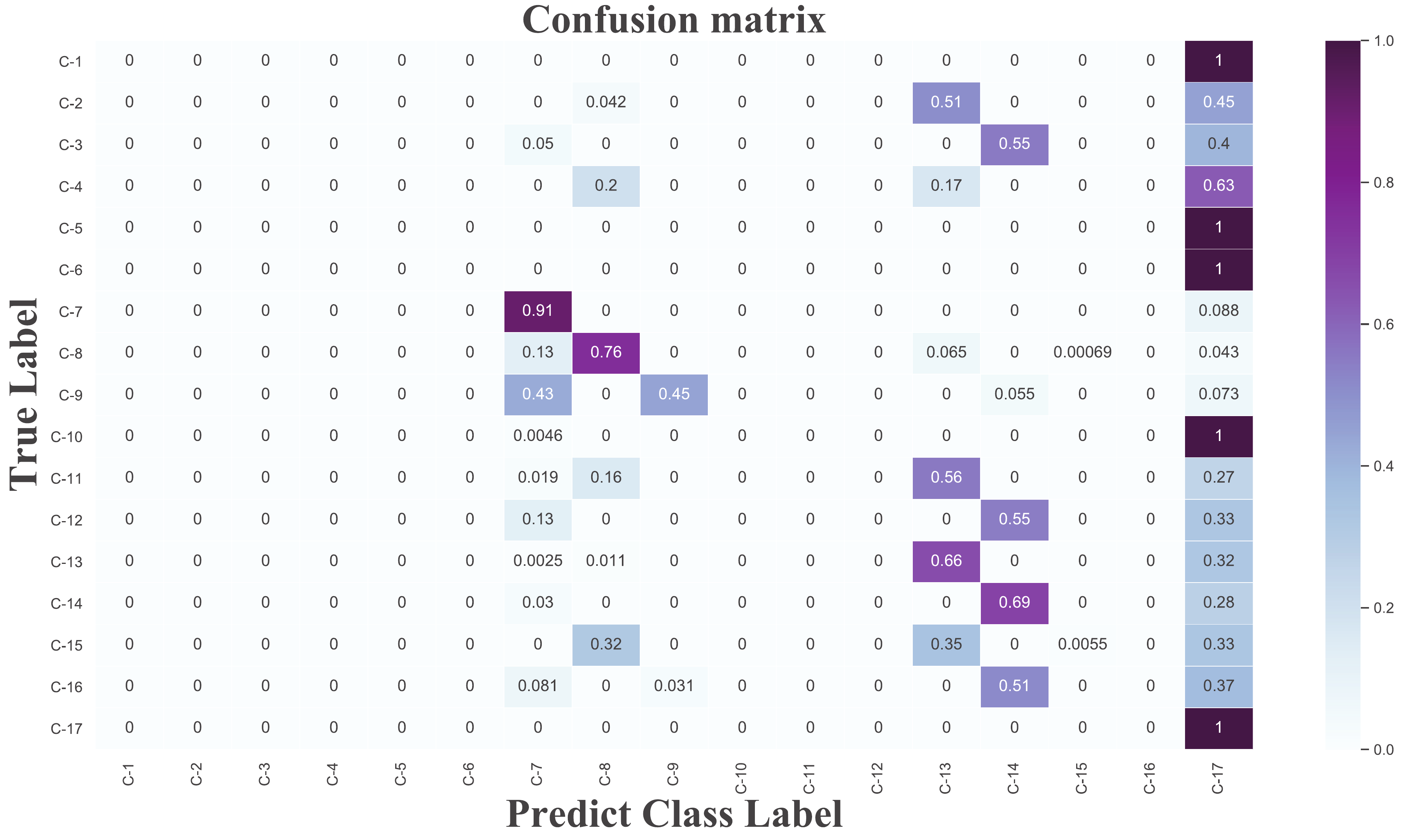}}
	\caption{A normalized confusion matrix  obtained using the SAA.}
	\label{Confusion_matrix_yes}
	\end{figure}

\begin{figure}[H]
		\centering
		\centerline{\includegraphics[scale=0.45]{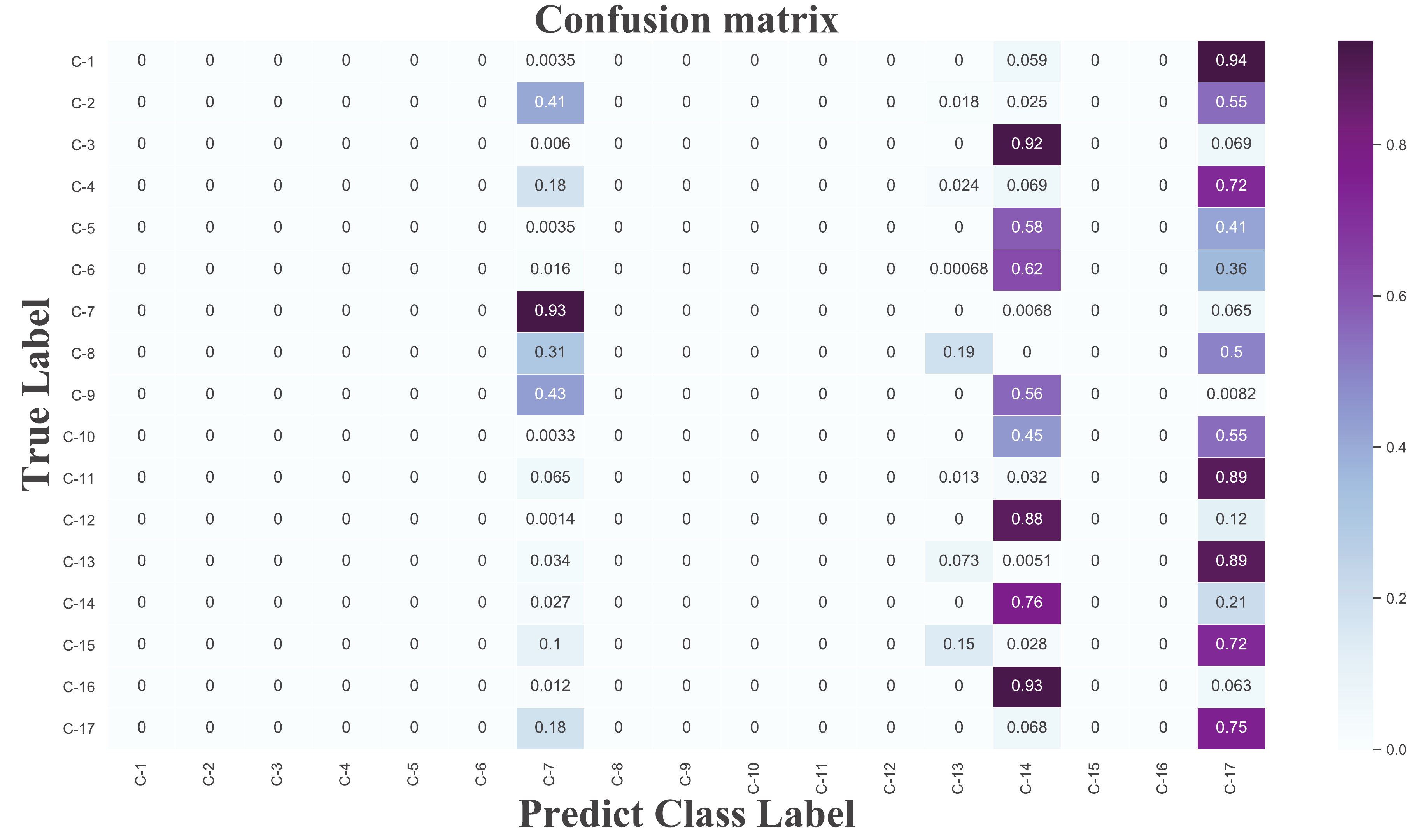}}
		\caption{A normalized confusion matrix  obtained using the FGSM.}
		\label{Confusion_matrix_yes_FGSM}
	\end{figure}

\begin{figure}[H]
		\centering
		\centerline{\includegraphics[scale=0.45]{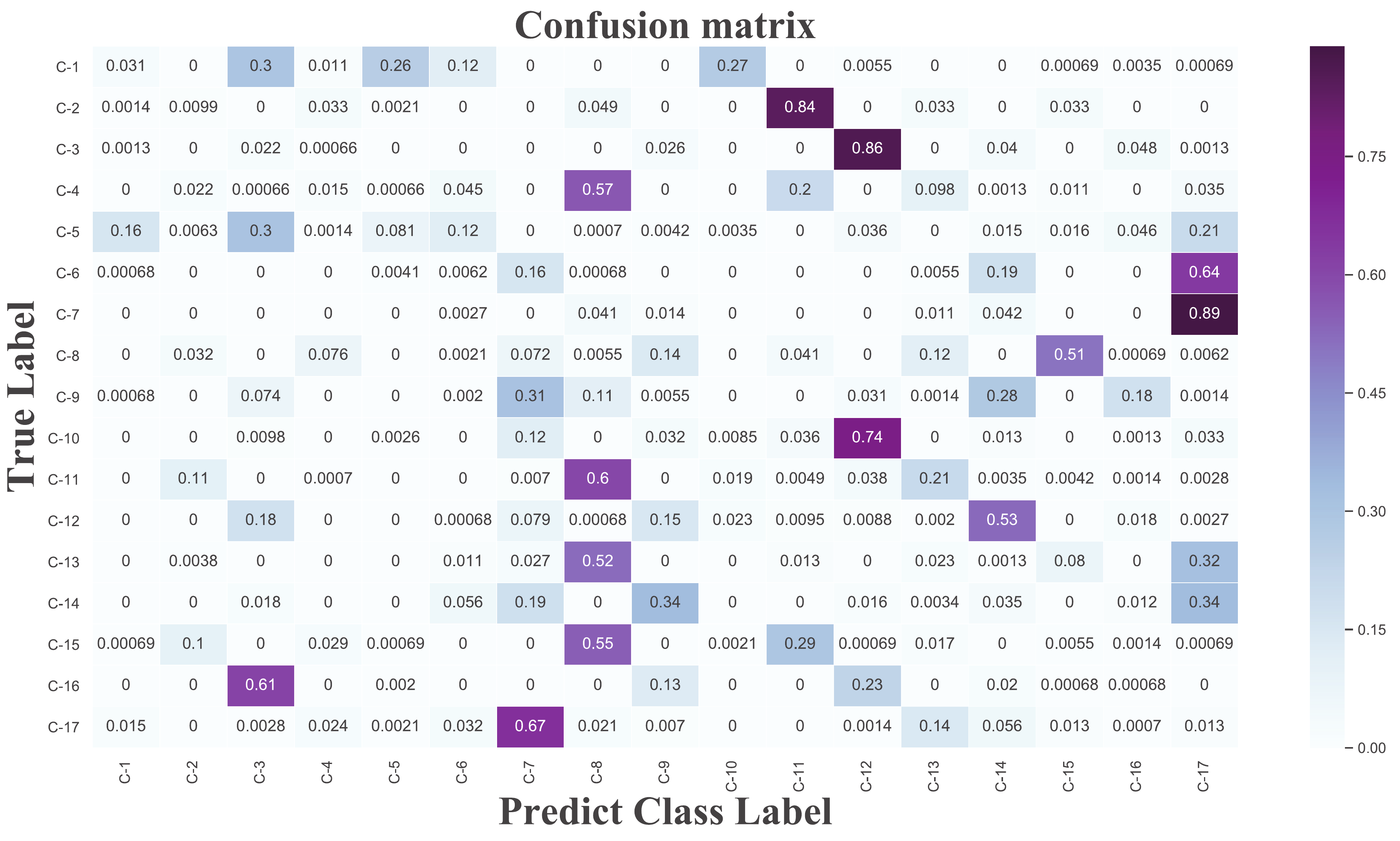}}
		\caption{A normalized confusion matrix  obtained using the SSA.}
		\label{Confusion_matrix_yes_SSA}
	\end{figure}

\begin{figure}[H]
		\centering
		\centerline{\includegraphics[scale=0.45]{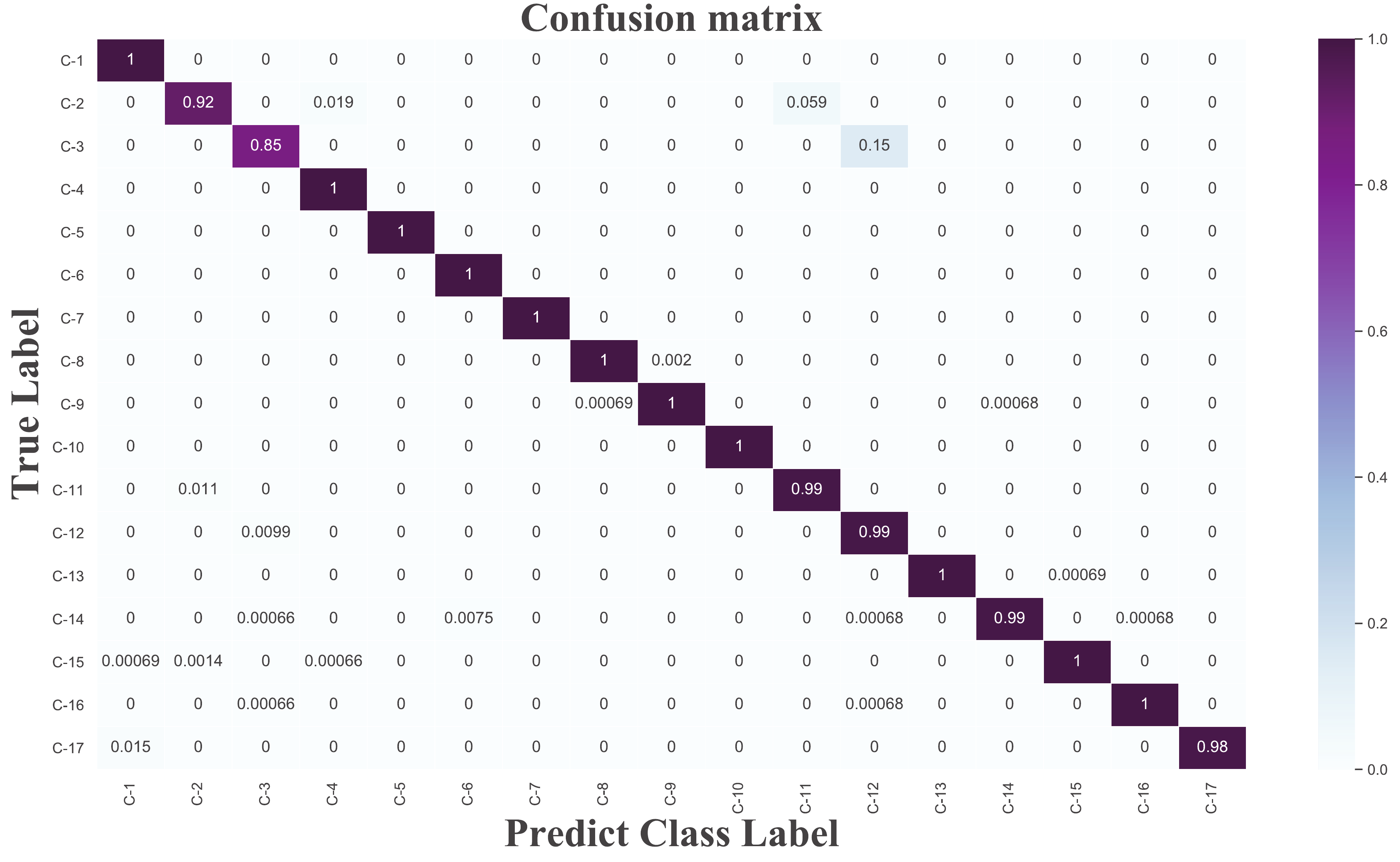}}
		\caption{A normalized confusion matrix obtained using the normal test signals.}
		\label{Confusion_matrix_yes_normal}
	\end{figure}

	We normalized confusion matrices generated according to
	class predictions obtained using SAA, FGSM and SSA in
	Fig.~\ref{Confusion_matrix_yes}, Fig.~\ref{Confusion_matrix_yes_FGSM} and Fig.~\ref{Confusion_matrix_yes_SSA}, respectively. For comparison, the normalized confusion matrix based on the normal test signals is also provided in Fig.~\ref{Confusion_matrix_yes_normal}. These normalized confusion
	matrices are used to construct class-confusion graphs
	proposed below.

	
	We analyze how class predictions of learned models change under attacks exploring class confusion statistics. We construct a directed graph ${G=\left( V,E \right)}$, where each node $v_i \in V$ denotes a class label C-$i$. A weighted directed edge $e_{i,j} \in E$ connecting a node $v_i$ to the node $v_j$ indicates that signals belonging to the class ${\rm C-}i$, are misclassified as ${\rm C-}j$. The weight $w _{i,j}$ of the edge $e_{i,j}$ is the ratio of the number of samples of the class ${\rm C-}i$ which are misclassified as ${\rm C-}j$, to the total number of samples in the class ${\rm C-}i$.

	\begin{figure*}
		\centering
		\begin{subfigure}[b]{.3\textwidth}
			\centering
			\includegraphics[scale=0.32]{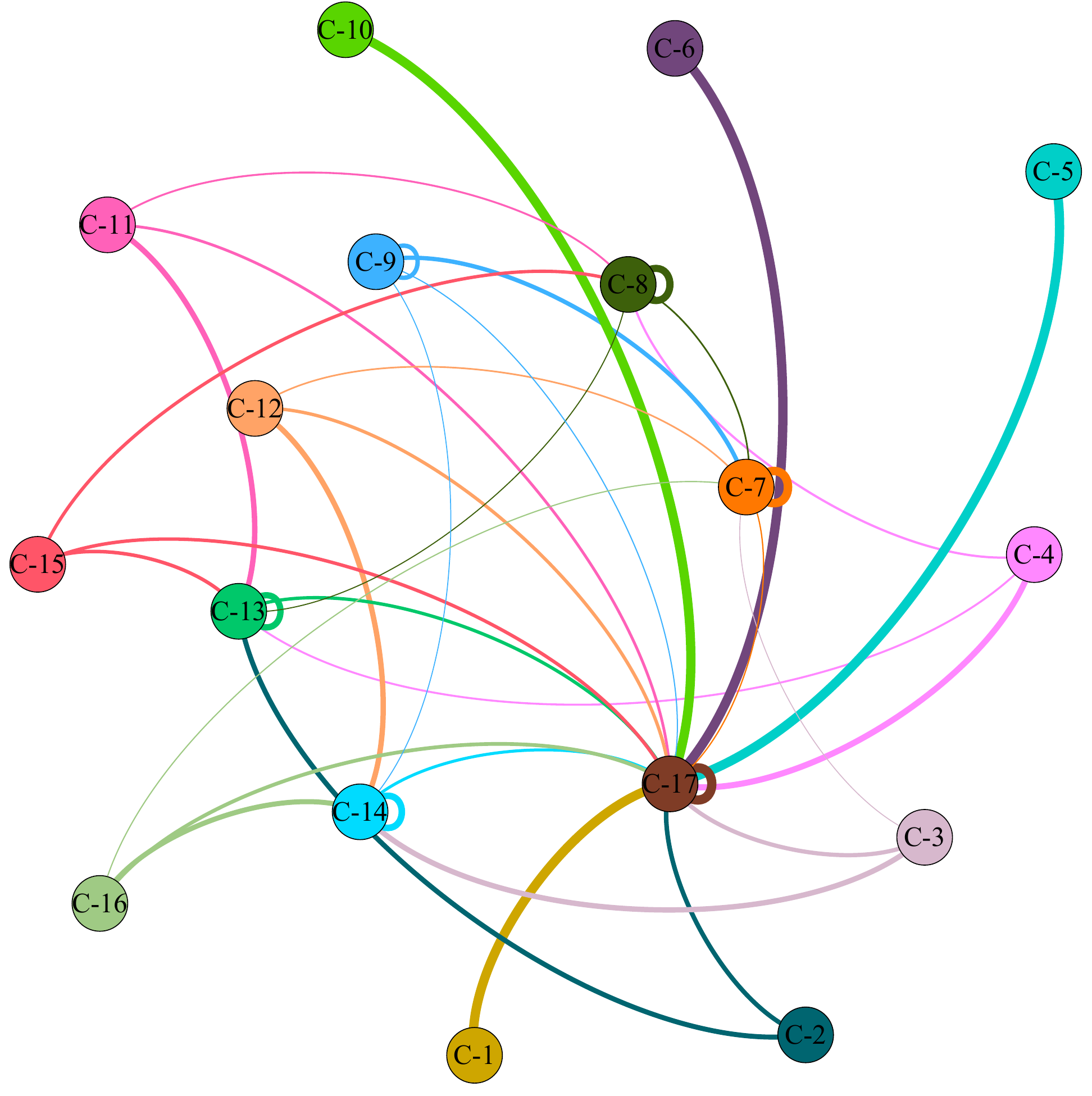}
			\caption{SAA}
			\label{fig:y equals x}
		\end{subfigure}
		\hfill
		\begin{subfigure}[b]{.3\textwidth}
			\centering
			\includegraphics[scale=0.32]{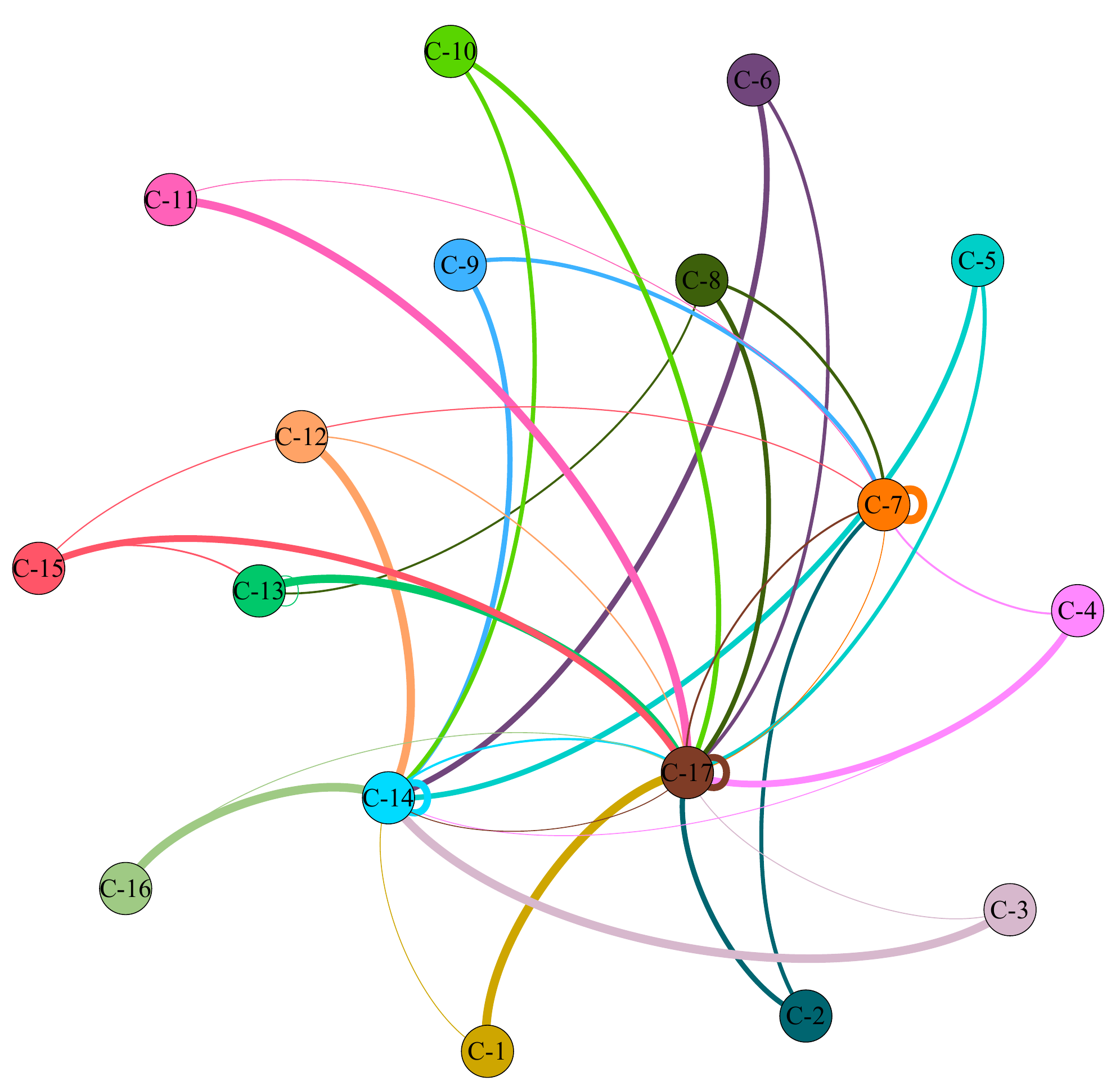}
			\caption{FGSM}
			\label{fig:three sin x}
		\end{subfigure}
		\hfill
		\begin{subfigure}[b]{.3\textwidth}
			\centering
			\includegraphics[scale=0.32]{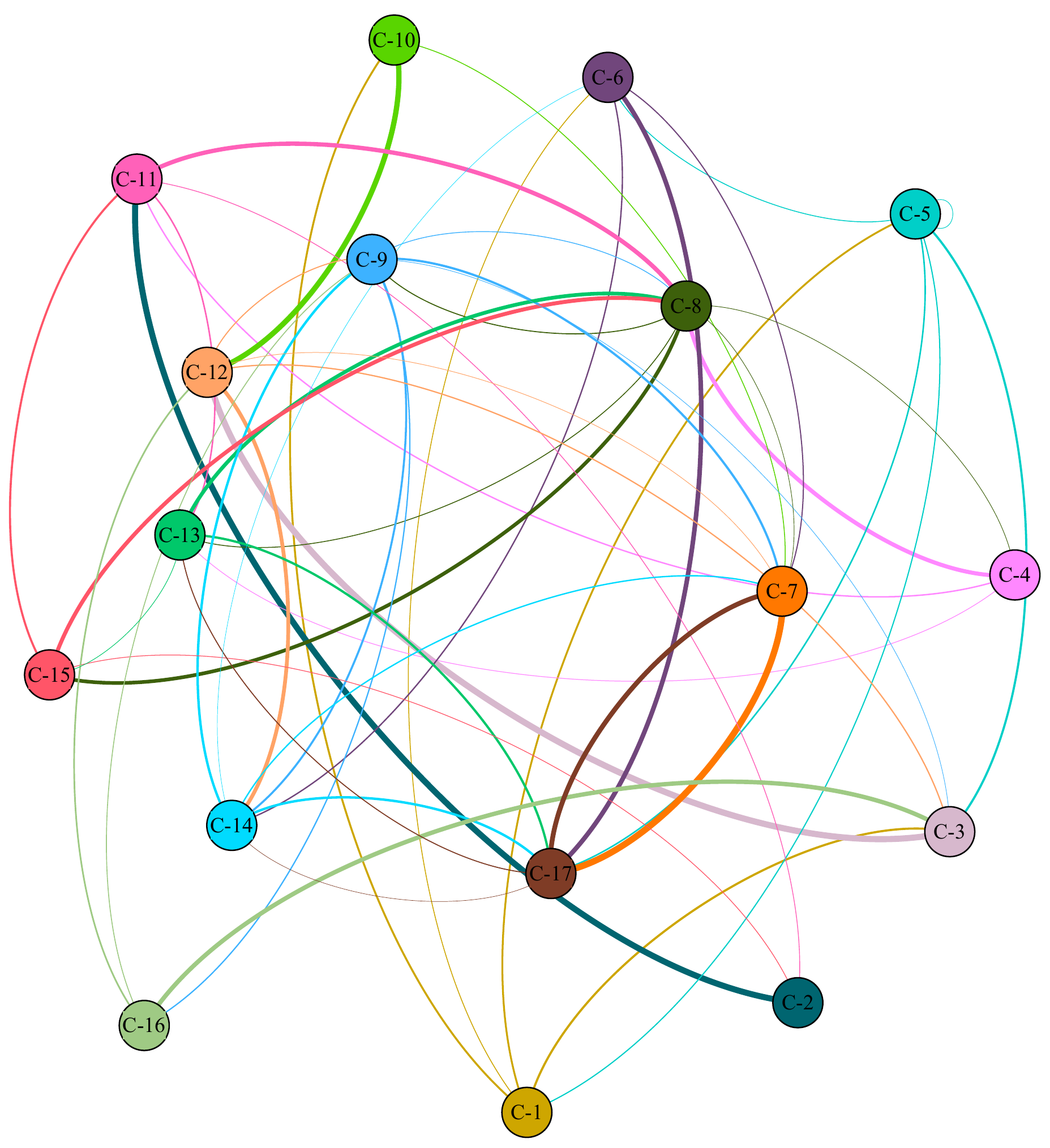}
			\caption{SSA}
			\label{fig:five over x}
		\end{subfigure}
		\caption{Visualization of graphs for each attack method. Directions of edges are depicted by color instead of using directed arrows. For instance, the edge {\color{ForestGreen}$e_{10, 17}$} is directed from the node {\color{ForestGreen}$C$-10} to the node {\color{brown}$C$-17}. Therefore, the color of the edge {\color{ForestGreen}$e_{10, 17}$} which is {\color{ForestGreen}green}, is same to that of the node {\color{ForestGreen}$C$-10} in  (a).  Thickness of edges is determined using normalized covariance matrices. For instance, in (a), the number of samples belonging to {\color{ForestGreen}$C$-10}  which are  misclassified as {\color{brown}$C$-17} is larger than that of samples belonging to {\color{ForestGreen}$C$-16}  which are misclassified as {\color{brown}$C$-17}.}
		\label{fig:three graphs}
	\end{figure*}

	\begin{table*}[t]
		\caption{Degrees of nodes of the class-confusion graphs of FGSM, SSA and SAA.}
		\centering
		\scalebox{1}{
			\begin{tabular}{|c|c|c|c|c|c|c|c|c|c|c|c|c|c|c|c|c|c|}
				\hline
				FGSM   & C-1 & C-2 & C-3 & C-4 & C-5 & C-6 & C-7 & C-8 & C-9 & C-10 & C-11 & C-12 & C-13 & C-14 & C-15 & C-16 & C-17   \\ \hline
				In-degree ($d_{in}$)  & 0   & 0   & 0   & 0   & 0   & 0   & 8   & 0   & 0   & 0    & 0    & 0    & 3    & 11   & 0    & 0    & 16    \\ \hline
				Out-degree ($d_{out}$) & 2   & 2   & 2   & 3   & 2   & 2   & 2   & 3   & 2   & 2    & 2    & 2    & 2    & 2    & 3    & 2    & 3     \\ \hline \hline
				
				SSA   & C-1 & C-2 & C-3 & C-4 & C-5 & C-6 & C-7 & C-8 & C-9 & C-10 & C-11 & C-12 & C-13 & C-14 & C-15 & C-16 & C-17   \\ \hline
				In-degree ($d_{in}$) & 1   & 2   & 5   & 1   & 2   & 3   & 7   & 5   & 4   & 1    & 3    & 3    & 4    & 4    & 2    & 1    & 5     \\ \hline
				Out-degree ($d_{out}$) & 4   & 1   & 1   & 3   & 5   & 3   & 1   & 5   & 5   & 2    & 3    & 4    & 3    & 4    & 3    & 3    & 3    \\ \hline \hline
				
				SAA& C-1 & C-2 & C-3 & C-4 & C-5 & C-6 & C-7 & C-8 & C-9 & C-10 & C-11 & C-12 & C-13 & C-14 & C-15 & C-16 & C-17   \\ \hline
				In-degree ($d_{in}$) & 0   & 0   & 0   & 0   & 0   & 0   & 6   & 4   & 1   & 0    & 0    & 0    & 6    & 5    & 0    & 0    & 16    \\ \hline
				Out-degree ($d_{out}$) & 1   & 2   & 3   & 3   & 1   & 1   & 2   & 3   & 4   & 1    & 3    & 3    & 2    & 2    & 3    & 3    & 1     \\ \hline
		\end{tabular}}
		\label{deg_table}
	\end{table*}
	
	Class-confusion graphs computed for each attack method are visualized in  Fig.~\ref{fig:three graphs}. For the sake of clearance of visualizations,
	we only depict connections denoting classification probability greater than chance\footnote{There are 17 types of PQ signals and the number of signals of each type is the same. Therefore, we assume that the probability of each signal being classified into any kind of signal is $1/17$ (the assumption is only to better show corresponding class-confusion graphs, because displaying large number of edges in figures makes corresponding graphs look messy)., which is $1/17$.} Distributions of node degrees are given in Table~\ref{deg_table}. Out-degree value of a node $v_i$ associated with a class C$-i$ denotes the number classes to which samples of the C$-i$ are incorrectly assigned by misclassification. For instance, samples belonging to C$-15$ which are attacked by SSA are misclassified to the other three different classes C$-2$, C$-8$ and C$-11$  (see Fig.~\ref{fig:three graphs}). In-degree value of a node $v_i$ denotes the number classes whose samples are misclassified as C$-i$. For instance, samples belonging to C$-8$ and C$-13$ are misclassified to C$-15$ (see Fig.~\ref{fig:three graphs}).
	
	Fig.~\ref{fig:three graphs}(a) and Fig.~\ref{fig:three graphs}(b) show that perturbations generated by the FGSM and SAA make learned models misclassify signals to a few dominant classes such as $C$-7, $C$-13$, C$-14 and {$C$-17}. We conjecture that signals belonging to these dominant classes are distributed in larger areas in feature spaces (see Fig.~\ref{all_visual}), and thereby, models can be fooled by assigning labels of signals to these classes \textit{with higher probability} compared to the other classes. Besides, most of the signals belonging to $C$-17, are not misclassified when universal perturbations are applied.
	The reason may be that decision boundary of $C$-17 may be more complicated compared to that of the other classes, and located far from that of the other classes (see Fig.~\ref{all_visual}). Therefore, universal perturbations may not be able to fool models for misclassifying signals of $C$-17. 
	
	In order to analyze the statistical relationship between classes whose samples are successfully attacked and failed to be attacked, we compute entropy of each node $v_i$ by
	\begin{equation}
	H_w(v_i) \triangleq -\sum_{j=1}^{d_i}{p_{i,j}\log ( p_{i,j})}, \;  {\rm and} \;  p_{i,j}=\frac{w_{i,j}}{\sum\limits_{j=1}^{d_i}{w _{i,j}}}, 
	\end{equation}
	where $d_i \in \{ d_{in}, d_{out} \}$ is either in-degree or out-degree of $v_i$. We compute the entropy for all the nodes of the graph $G$ by
	\begin{equation}
	H_w(G) \triangleq \sum_{v_i \in V} H_w(v_i).
	\end{equation}
	
	We calculated unweighted graph entropy $H(G)$ using ${w_{i,j}=1}, \forall e_{i,j} \in E$. Entropy values  given in Table~\ref{entropy} show that SSA provides the largest entropy values. This result shows that signals misclassified under SSA are more equally likely distributed to incorrect classes compared to FGSM and SAA. We obtained the lowest values of $H_{w}(G)$ and $H(G)$ for the in-degree  distribution, and those of $H_{w}(G)$ for the out-degree distribution using  FGSM. However, SAA provided the lowest unweighted entropy for the out-degree distribution. This result indicates that SAA makes the learned model misclassify samples to less number of particular signal classes with higher probability compared to  FGSM and SAA. This observation is depicted for the graph computed using the SAA with less number of node hubs in Fig.~\ref{fig:three graphs}(a), compared to the graphs computed using the FGSM (Fig.~\ref{fig:three graphs}(b)) and the SSA (Fig.~\ref{fig:three graphs}(c)). 


	\begin{table}[t]
		\caption{Weighted and unweighted graph entropy values.}
		\centering
		\scalebox{1.20}{
			\begin{tabular}{|c|c|c|c|c|c|c|}
				\hline
				Entropy                              & \multicolumn{3}{c|}{Out-degree ($d_{out}$)} & \multicolumn{3}{c|}{In-degree ($d_{in}$)} \\ \hline
				\multirow{2}{*}{$H_w(G)$}   & FGSM      & SSA      & SAA      & FGSM     & SSA      & SAA      \\ \cline{2-7} 
				& 13.10  & 21.29 & 13.97 & 10.87 & 20.84 & 11.75 \\ \hline
				\multirow{2}{*}{$H(G)$} & FGSM      & SSA      & SAA      & FGSM     & SSA      & SAA      \\ \cline{2-7} 
				& 19.34  & 25.06 & 17.09 & 12.04 & 23.53 & 13.49 \\ \hline
		\end{tabular}}
		\label{entropy}
	\end{table}

	\begin{table*}[ht]
		\caption{Comparison of classification accuracy ($\%$) for adversarial training (adv.) using SSA.}
		\centering
		\scalebox{1.10}{\begin{tabular}{|c|c|c|c|c|}
				\hline
				Adv. Training Epochs & Acc. on Training Data &	Acc. on Test Data & Acc. on Adv. Training Data & Acc. on Adv. Test Data \\ \hline
				0 & 98.54& 98.46 & 2.09 & 2.09            \\ \hline
				1 & 91.31 $\pm$ 10.03 & 91.08 $\pm$ 11.33 & 90.14 $\pm$ 10.14 & 90.06 $\pm$ 10.12          \\ \hline
				10 & 83.26 $\pm$ 7.18 & 83.14 $\pm$ 8.29 & 93.15 $\pm$ 5.72 & 93.26 $\pm$ 5.52          \\ \hline
		\end{tabular}}
		\label{adv}
	\end{table*}



	
	\subsection{Analysis of Performance of Adversarial Training}
	We fine-tuned the learned model for 10 additional epochs using generated adversarial signals to improve its robustness against adversarial perturbation. Although we generated adversarial signals for training  and test data, only adversarial signals in training data were used for adversarial training.  Adversarial signals in  test data were used to evaluate performance of adversarial training. For each trial, same adversarial signals were used in all additional epochs. We fine-tuned the model for 10 more epochs using the original data for a fair comparison. 
	
\begin{figure}[t]
		\centerline{\includegraphics[scale=0.5]{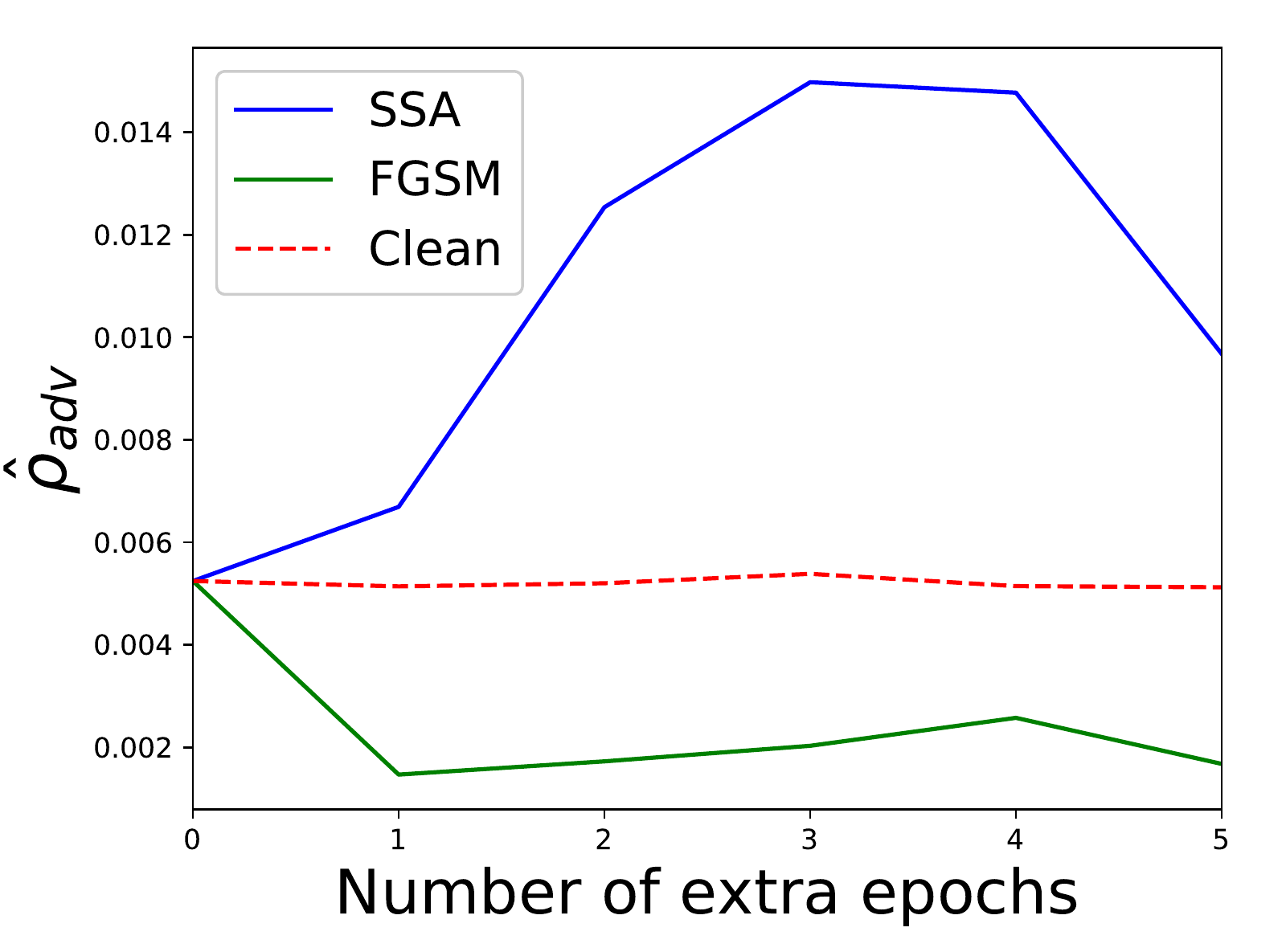}}
		\caption{Analysis of change of average robustness $\hat{\rho}_{\mathrm{adv}}$ of the learned models using SSA and FGSM.}
		\label{robustness_advtrain_epoch}
	\end{figure}

	Change of $\hat{\rho}_{\mathrm{adv}}$ for different adversarial training strategies is depicted in Fig.~\ref{robustness_advtrain_epoch}. The robustness $\hat{\rho}_{\mathrm{adv}}$ is estimated using SSA, since it is the best performing method, while worst performing SAA is not used in this experiment (see Table~\ref{FGSM vs SSA}).  Fig.~\ref{robustness_advtrain_epoch} shows that adversarial training using adversarial signals generated by SSA significantly increases robustness of the learned models to attacks. For example, robustness of the learned models is improved by 200$\%$ after training by three epochs. However, training the learned models with adversarial signals generated by FGSM \cite{B14} can decrease the robustness of the learned models (note that we used relatively large perturbations by FGSM $\epsilon=0.5$ due to its high  misclassification rate). The reason of this result may be that the perturbations generated by the FGSM are larger than those generated by the SSA. Adversarial training with extremely perturbed signals decreases robustness of the learned models to adversarial perturbations \cite{B5}. 
	
	
	\subsection{Accuracy-Robustness Trade-off}
	
	Next, learned models are fine-tuned using combined signals (mixing of normal signals and adversarial signals generated by SSA) to analyze change of their accuracy (FGSM and SAA are not used since they need relatively large perturbations to achieve high misclassification rate). Results show that although adversarial training improved the classification accuracy using adversarial PQ signals to $94.09\%$, the classification accuracy of the  model retrained using normal PQ signals is decreased to $81.34\%$ after training with 10 additional epochs (see Table~\ref{adv}). The reason may be that fine-tuning of models with generated adversarial signals can decrease the model's ability to express normal signals to a certain extent. This phenomenon also exists in other fields such as the image domain \cite{zhang2019theoretically,arani2020adversarial}. We propose the trade-off between maintaining accuracy of classifiers on normal PQ signals and improving their robustness to adversarial signals as an open challenge, which needs to be better utilized and resolved in future research.
	
	\subsection{Visualization of Learned Feature Representations of PQ Signals under Attacks}
	Visual inspection of attacks on PQ signals in the Smart Grid is not a well-defined phenomenon, since their interpretation is not clear using  measurement devices. To address this problem, we propose an approach by first projecting high dimensional signal and feature vectors of  samples  to low dimensional spaces. Then, we visualize samples in projected spaces to explore their discrimination among PQ classes. 
	
	\begin{figure}[t]
		\centerline{\includegraphics[scale=0.5]{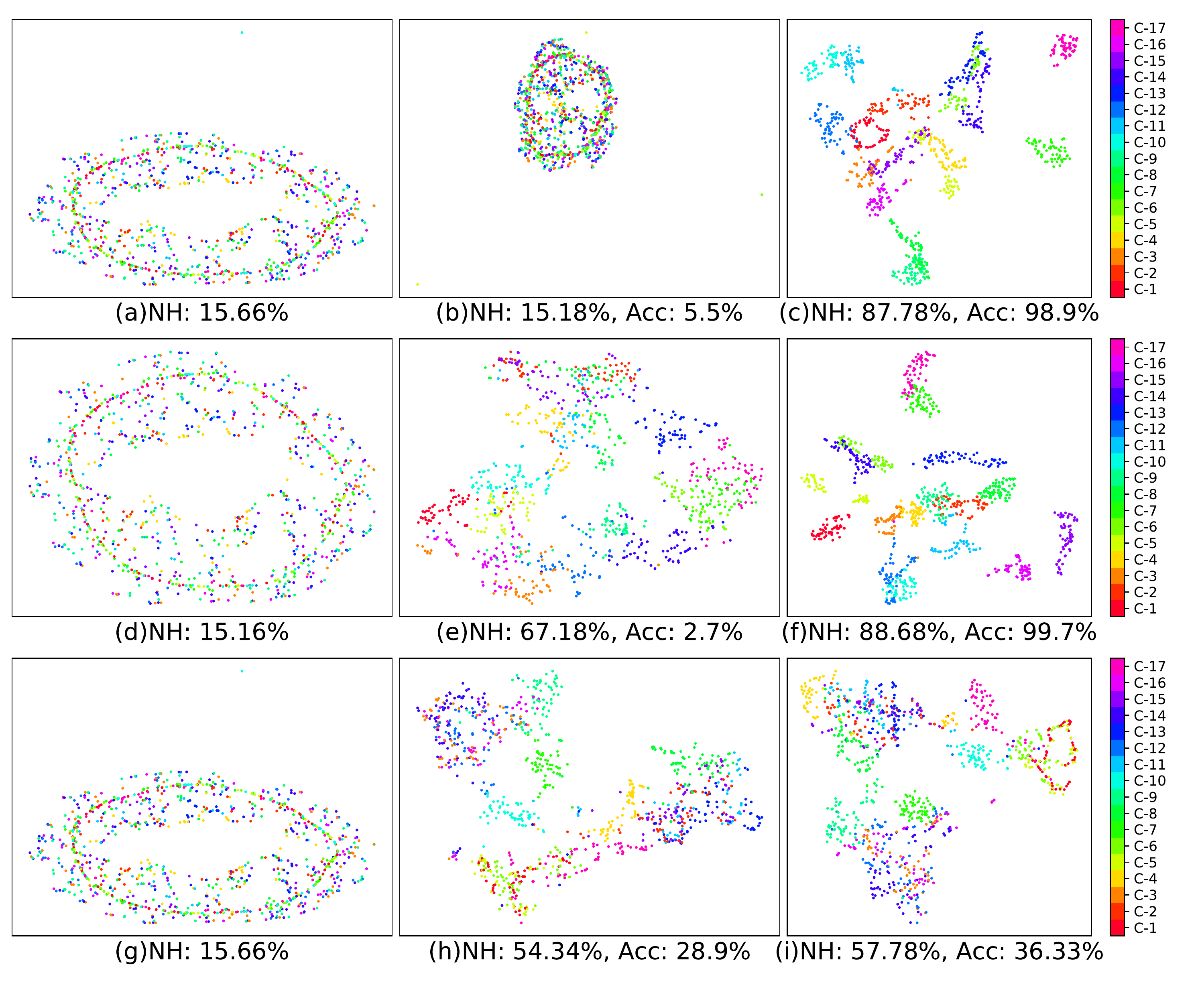}}
		\caption{Visualization of the PQ signals in (a) $\mathcal{X}_n$, (b) $\mathcal{F}_{n}^{r}$, (c)~$\mathcal{F}_n^{tr}$, (d) $\mathcal{X}_{ssa}$, (e) $\mathcal{F}_{ssa}^{tr}$, (f) $\mathcal{F}_{ssa}^{adv}$, (g) $\mathcal{X}_{saa}$, (h) $\mathcal{F}_{saa}^{tr}$, and (i) $\mathcal{F}_{saa}^{adv}$, with NH and classification accuracy (Acc).}
		\label{all_visual}
	\end{figure}

	We used t-SNE \cite{maaten2008visualizing} for projection of high dimensional vectors. We first projected signal vectors  residing in input signal spaces and feature vectors obtained at the last hidden layer of CNN models into a 2D space $\mathcal{X}$ and a 2D feature space $\mathcal{F}$, respectively. Then, we depicted samples on the projected spaces by points which are colored according to their class labels. Projection quality is evaluated by neighborhood hit (NH), which indicates how well samples belonging to classes are visually discriminated. For a given $k$ (in our work, $k = 5$), the NH for a point $x$ is the ratio
	of its $k$-nearest neighbors that belong to the same category as $x$ \cite{rauber2016visualizing}. The
	NH for the entire projection is the average NH over its points. 1000 randomly chosen normal PQ signals, adversarial signals generated by SSA and SAA are visualized in Fig.~\ref{all_visual}. Fig.~\ref{all_visual}(a), (d) and (g) show poor discrimination of normal samples, and adversarial signals generated by SSA and SAA in the projected spaces $\mathcal{X}_{n}$, $\mathcal{X}_{ssa}$ and $\mathcal{X}_{saa}$, respectively. This result shows difficulty of discriminating samples using the input signals (normal signals or adversarial signals). This qualitative result is also confirmed by their low NH values. 
	
	In Fig.~\ref{all_visual}(b), samples are depicted in a 2D feature space $\mathcal{F}_{n}^{r}$ obtained before training of CNN models. This result indicates that a CNN model with randomly initialized parameters has a poor visual discrimination. It is reasonable to hypothesize that visual discrimination would be improved \textit{after} training of models. This is related to the  conjecture that CNNs can learn features useful for human perception of classes \cite{zhang2018perceptual}. To analyze this hypothesis, we first trained a CNN model and extracted features of samples at the  last hidden layer of the model. Then, we projected features into a 2D feature space denoted by $\mathcal{F}_n^{tr}$. Samples are depicted in the projected space $\mathcal{F}_n^{tr}$ in Fig.~\ref{all_visual}(c). Compared to Fig.~\ref{all_visual}(a) and (b), visual discrimination is significantly improved in Fig.~\ref{all_visual}(c) and the corresponding NH is 87.78$\%$ and the test accuracy is 98.9$\%$. 
	
	Fig.~\ref{all_visual}(e) shows that features of adversarial signals generated by SSA are not visually better discriminative compared with Fig.~\ref{all_visual}(c). Therefore, the adversarial attacks affect visual discrimination of the features. The test accuracy for the 1000 randomly chosen adversarial signals generated by SSA is only 2.7$\%$. However, features of the adversarial signals are visually distinguishable to a certain degree, since the SSA can successfully fool top layer softmax classifiers for misclassification of samples.
	Visual discrimination of samples is improved in a projected space  $\mathcal{F}_{ssa}^{adv}$ of features learned \textit{after} adversarial training of the model (see Fig.~\ref{all_visual}(f)). This result visually verifies the efficiency of adversarial training. 
	
	In Fig.~\ref{all_visual}(i), we visualize samples generated by SAA in a projected space of features $\mathcal{F}_{saa}^{adv}$ learned \textit{after} adversarial training of the model. Although the NH is only 57.78$\%$, visual discrimination of adversarial PQ signals is improved by adversarial training. This may suggest that adversarial training of models using adversarial PQ signals generated by an attack method can improve their robustness against adversarial PQ signals generated by other attack methods. Results obtained using our proposed approach are complementary to those obtained by visualization of raw signals in Fig.~\ref{Universal}. We believe that researchers can use the proposed approach for qualitative analysis of class discrimination of normal and adversarial PQ signals under attacks in the Smart Grid.

	\section{Conclusion and Discussion}
	
	We proposed signal-specific adversarial attack (SSA) and signal-agnostic adversarial attack (SAA) methods to attack learning models for classification of power quality signals by generating adversarial signals in the Smart Grid. We proposed and evaluated black-box attacks based on transferable characteristics and the above two methods. We adopted adversarial training to defend learned models against adversarial signals. Experimental results show that our proposed SSA provides less perturbation compared to the state-of-the-art FGSM which was adapted for power signal analysis, while the SAA can generate the universal perturbation that can fool learned models. The attack method based on the universal signal-agnostic algorithm has a higher transfer rate of black-box attacks than the attack method based on the signal-specific algorithm, which is a huge
and practical threat for CNN based PQ classification. In the analyses, the proposed adversarial training method improved robustness of the learning models. In addition, we proposed an approach for qualitative analysis of normal power quality signals and visual inspection of adversarial attacks. We elucidated the mechanisms behind robustness of models to adversarial attacks corresponding to the quantitative results using the proposed visualization approach. We believe that the proposed approaches will be useful for researchers and practitioners to analyze power systems and lead new research directions. In the future work, we plan to employ our approaches for a wider class of attack models (such as gray-box and black-box attacks with higher misclassification rates) and defense methods for data, network and device security in the Smart Grid.

	
	%




	\ifCLASSOPTIONcaptionsoff
	\newpage
	\fi

	
	
	%
	\bibliographystyle{IEEEtran}
	\bibliography{IEEEabrv,mybibfile}

	%

\end{document}